\documentclass[twocolumn,prl,floatfix,superscriptaddress,longbibliography]{revtex4-2}
\usepackage[hidelinks,colorlinks=true,citecolor=blue]{hyperref}

\usepackage{float}
\usepackage[dvipsnames]{xcolor}
\usepackage{graphicx}
\usepackage{amsmath}
\usepackage{bm}
\usepackage{ulem}
\usepackage{stix}

\begin{document}

\title{Superconductivity governed by Janus-faced fermiology in strained bilayer nickelates}

\author{Siheon Ryee}
 \email{siheonryee@gmail.com} 
\affiliation{I. Institute of Theoretical Physics, University of Hamburg, Notkestra{\ss}e 9-11, 22607 Hamburg, Germany}
\affiliation{The Hamburg Centre for Ultrafast Imaging, Luruper Chaussee 149, 22761 Hamburg, Germany}

\author{Niklas Witt}
\affiliation{Institut f\"ur Theoretische Physik und Astrophysik and W\"urzburg-Dresden Cluster of Excellence ct.qmat, Universit\"at W\"urzburg, 97074 W\"urzburg, Germany}

\author{Giorgio Sangiovanni}
\affiliation{Institut f\"ur Theoretische Physik und Astrophysik and W\"urzburg-Dresden Cluster of Excellence ct.qmat, Universit\"at W\"urzburg, 97074 W\"urzburg, Germany}

\author{Tim O. Wehling}
\affiliation{I. Institute of Theoretical Physics, University of Hamburg, Notkestra{\ss}e 9-11, 22607 Hamburg, Germany}
\affiliation{The Hamburg Centre for Ultrafast Imaging, Luruper Chaussee 149, 22761 Hamburg, Germany}

\date{\today}

\begin{abstract}
High-temperature superconductivity in pressurized and strained bilayer nickelates (La,Pr)$_3$Ni$_2$O$_7$ has emerged as a new frontier. One of the key unresolved issues concerns the fermiology that underlies superconductivity. On both theoretical and experimental sides, no general consensus has been reached, and conflicting results exist regarding whether the relevant Fermi surface involves a $\gamma$ pocket---a hole pocket with $d_{z^2}$-orbital character centered at the Brillouin zone corner. Here, we address this issue by unveiling a Janus-faced role of the $\gamma$ pocket in spin-fluctuation-mediated superconductivity. We show that this pocket simultaneously induces dominant pair-breaking and pair-forming channels for the leading $s_\pm$-wave pairing. Consequently, an optimal superconducting transition temperature $T_\mathrm{c}$ is achieved when the $\gamma$ pocket surfaces at the Fermi level, placing the system near a Lifshitz transition. This suggests that superconductivity can emerge, provided the maximum energy level of the $\gamma$ pocket lies sufficiently close to the Fermi level, either from below or above. Our finding not only reconciles two opposing viewpoints on the fermiology, but also naturally explains recent experiments on (La,Pr)$_3$Ni$_2$O$_7$ thin films, including the superconductivity under compressive strain, two conflicting measurements on the Fermi surface, and the dome shape of $T_\mathrm{c}$ as a function of hole doping.
\end{abstract}
 
\maketitle

\begin{figure} [!htbp] 
	\includegraphics[width=1.0\columnwidth, angle=0]{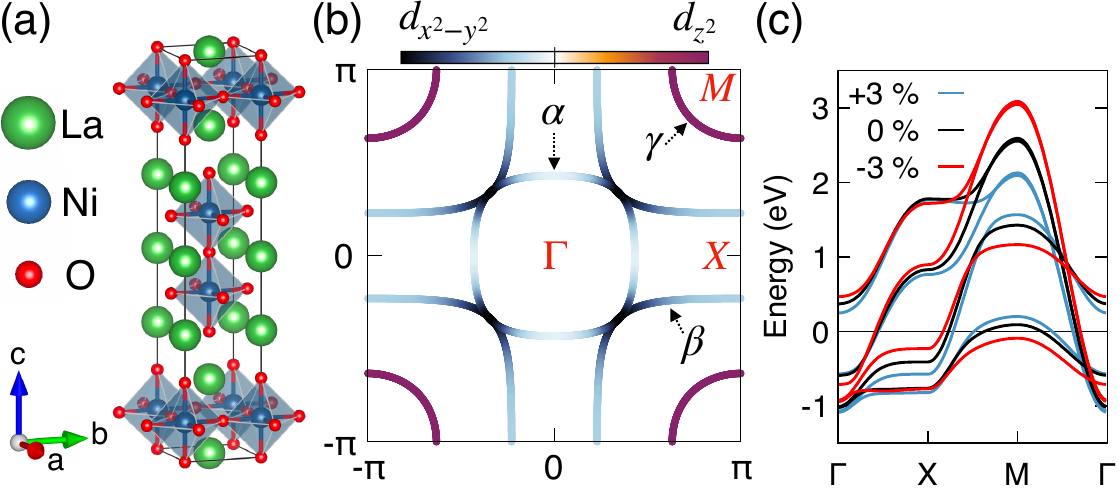}
	\caption{(a) Crystal structure of La$_3$Ni$_2$O$_7$ drawn using VESTA \cite{VESTA}. (b) Fermi surface obtained from the MLWF fit of the DFT band structure at zero strain. The color bar indicates the relative orbital weight. (c) The evolution of the band structure with strain. $+$ and $-$ denote tensile and compressive strain, respectively. 
	}
	\label{fig1}
\end{figure}

The discovery of high-temperature superconductivity in pressurized bilayer nickelate La$_3$Ni$_2$O$_7$ has opened a new frontier in superconductivity research \cite{sun_signatures_2023}. This material features a bilayer square lattice as a building block [Fig.~\ref{fig1}(a)], where three electrons are distributed across four Ni-$e_g$ orbitals in the top and bottom layers, while fully occupied Ni-$t_{2g}$ orbitals remain inert to low-energy physics~\cite{sun_signatures_2023,yang_orbital-dependent_2023,luo_bilayer_2023,gu_effective_2023,shen2023effective,christiansson_correlated_2023,zhang_trends_2023}. At ambient pressure, La$_3$Ni$_2$O$_7$ exhibits a (spin) density wave order below $140$~K, and the associated transition temperature is gradually decreased with increasing pressure~\cite{liu_evidence_2022,chen2023evidence,chen2024electronic,khasanov2025}. Above a critical pressure of $\sim$10~GPa, superconductivity sets in, reaching a maximum critical temperature $T^\mathrm{max}_\mathrm{c} \simeq 80$~K near 20~GPa~\cite{sun_signatures_2023,hou_emergence_2023,zhang_high-temperature_2023,wang_pressure-induced_2023,zhang_effects_2023,li2025_100gpa}. 
This pressure-induced superconductivity is also observed in its isostructural sibling compounds  La$_2$PrNi$_2$O$_7$~\cite{wang2024la2prni2o7} and  La$_2$SmNi$_2$O$_{7-\delta}$~\cite{li2025ambientpressuregrowthbilayer}, and related cousin systems including trilayer nickelate La$_4$Ni$_3$O$_{10}$~\cite{sakakibara_theoretical_2023,zhu2024la4ni3o10,li2024la4ni3010}, alternately stacked trilayer–monolayer polymorph of La$_3$Ni$_2$O$_7$~\cite{chen2024polymorph,puphal2024,abadi2025}, and La$_5$Ni$_3$O$_{11}$~\cite{shi2025superconductivityhybridruddlesdenpopperla5ni3o11}.

Despite extensive studies~\cite{luo_bilayer_2023,gu_effective_2023,shilenko_correlated_2023,christiansson_correlated_2023,zhang_trends_2023,sakakibara_possible_2023,oh_type_2023,lechermann_electronic_2023,liao_electron_2023,qin_high-t_c_2023,lu_interlayer_2023,tian_correlation_2023,yang_possible_2023,lu_superconductivity_2023,wu_charge_2023,labollita_electronic_2023,cao_flat_2023,qu_bilayer_2023,yang_strong_2023,yang2023valence,jiang2023high,sakakibara_theoretical_2023,zhang_structural_2023,huang2024impurity,shen2023effective,ryee2024_quenched,zhang2024strong,fan2024superconductivity,zhang2024prediction,zhang2024spm,yang2024effective,zhang2024alternating,yang2024decomposition,lu2024interplay,heier2024competing,sui2024electronic,geisler2024structural,rhodes2024,labollita2024_trilyaer,geisler2024optical,labollita2024_polymorph,lechermann2024,botzel2024,luo2024hightc,kakoi2024_pair,botzel2025theory,xi2025_dwave,zhan2025impact,xia2025_sensitive,shao2025bandstructurepairingnature,devaulx2025pressurestraineffectstextitab,lechermann2025lowenergy,nomura2025,maier2025interlayer,yue2025correlated}, no general consensus has been reached on fundamental aspects of superconductivity, including the pairing mechanism, gap symmetry, and underlying fermiology. A prevailing view is that strong magnetic interactions or fluctuations play a central role in the emergence of superconductivity in this multilayer nickelate family~\cite{zhang_structural_2023,sakakibara_theoretical_2023,zhang_trends_2023,sakakibara_possible_2023,oh_type_2023,lechermann_electronic_2023,liao_electron_2023,qin_high-t_c_2023,lu_interlayer_2023,tian_correlation_2023,yang_possible_2023,lu_superconductivity_2023,lu_interlayer_2023,tian_correlation_2023,yang_possible_2023,lu_superconductivity_2023,wu_charge_2023,qu_bilayer_2023,yang_strong_2023,yang2023valence,jiang2023high,ryee2024_quenched,shen2023effective,zhang2024strong,fan2024superconductivity,zhang2024prediction,zhang2024spm,yang2024effective,zhang2024alternating,yang2024decomposition,lu2024interplay,heier2024competing,luo2024hightc,kakoi2024_pair,botzel2025theory,xi2025_dwave,zhan2025impact,xia2025_sensitive,le2025landscape,yue2025correlated}.
In this line of thought, there is an issue of whether or not the ``flat'' $\gamma$ Fermi surface (FS) pocket centered at the $M$ point of the fist Brillouin zone (FBZ) [Fig.~\ref{fig1}(b)] exists in the normal state. This question is deeply linked to whether the $\gamma$ pocket is the main driver of the pairing fluctuations within the spin-fluctuation theory. Although the common belief is that the $\gamma$ pocket is present and mediates pairing~\cite{sun_signatures_2023,zhang_trends_2023,lechermann_electronic_2023,yang_possible_2023,lu_superconductivity_2023,lu_interlayer_2023,tian_correlation_2023,yang_possible_2023,jiang2023high,fan2024superconductivity,zhang2024prediction,zhang2024spm,yang2024effective,zhang2024alternating,heier2024competing,xi2025_dwave,zhan2025impact,le2025landscape,yue2025correlated}, there are different perspectives challenging this viewpoint~\cite{ryee2024_quenched,christiansson_correlated_2023,devaulx2025pressurestraineffectstextitab,dessau2017}. This controversy over ``three vs two FS pockets'' remains unresolved, partly because high pressure limits the use of many experimental probes.

In this respect, the recent discovery of superconductivity in thin films of La$_3$Ni$_2$O$_7$~\cite{ko2025thinfilm} and La$_2$PrNi$_2$O$_7$~\cite{zhou2025thinfilm,liu2025thinfilm} under ambient pressure offers a promising avenue for resolving the issues by enabling application of a broad range of experimental techniques. The key finding is that superconductivity with  $T_\mathrm{c} \simeq 42$~K or higher emerges in compressively strained samples ($- 2$~\%  strained thin films on SrLaAlO$_4$ substrate~\cite{ko2025thinfilm,zhou2025thinfilm,liu2025thinfilm}). These are striking observations, because the standard {\it ab initio} calculations based on density functional theory (DFT) unequivocally point out that compressive strain suppresses the $\gamma$ pocket~\cite{zhao2025_strain,geisler2025compressive,yi2025unifying}; see Fig.~\ref{fig1}(c). Interestingly, recent angle-resolved photoemission spectroscopy (ARPES) measurements have reported conflicting results regarding the presence of this pocket~\cite{li2025arpes, wang2025arpes}. Taken together, these findings reignite the central issue concerning the presence of the $\gamma$ pocket and its role in spin-fluctuation-mediated superconductivity.

In this work, we solve this problem by identifying a {\it Janus-faced} role of the $\gamma$ pocket in the  superconductivity of strained bilayer nickelate (La,Pr)$_3$Ni$_2$O$_7$. Namely, the $\gamma$ pocket simultaneously
generates both strong pair-forming and pair-breaking channels, so a delicate balance between them is the key to superconductivity. As a consequence, maximum $T_\mathrm{c}$ can be achieved when the $\gamma$ pocket has just surfaced at the Fermi level, thereby being in proximity to a Lifshitz transition. Our finding not only reconciles two seemingly contradicting perspectives on the fermiology (i.e., three vs two pockets), but also accounts for experimental observations in strained samples, including the enhancement of $T_\mathrm{c}$ with increased compressive strain~\cite{ko2025thinfilm,bhatt2025}, the presumably $s_\pm$-wave gap symmetry~\cite{fan2025gapsymmetry}, conflicting ARPES measurements on the Fermi surface \cite{li2025arpes,wang2025arpes}, and the dome shape of $T_\mathrm{c}$ as a function of hole doping~\cite{hao2025superconductivityphasediagramsrdoped}.

We first perform DFT calculations and then use the maximally localized Wannier function (MLWF) method \cite{marzari2012} to construct a four-orbital (top-layer two Ni-$e_g$ + bottom-layer two Ni-$e_g$) bilayer square-lattice model of strained La$_3$Ni$_2$O$_7$ using DFT-optimized lattice constants reported in Ref.~\cite{zhao2025_strain}. 
Since in-plane lattice constants only differ by about $1~\%$ in the orthorhombic Fmmm structure~\cite{zhao2025_strain,bhatt2025}, we use higher-symmetry tetragonal I4/mmm structure by averaging the in-plane lattice constants. 
Atomic positions are relaxed for each lattice constant using the Perdew–Burke–Ernzerhof (PBE) exchange-correlation functional for the generalized gradient approximation \cite{PBE}. We used the Vienna Ab initio Simulation Package (VASP)~\cite{Kresse1993, Kresse1996, Kresse1996b} for DFT calculations and Wannier90 code~\cite{Wanner90} for constructing MLWFs; see Supplementary for details~\cite{supple}. The FS and band structures in Figs.~\ref{fig1}(b--c) were obtained following this procedure.

To investigate the superconducting phase transition, we use the random phase approximation (RPA) and solve the gap equation given by 
\begin{align}
\begin{split}
	\lambda_{\mathrm{sc}} \Delta_{l_1 l_2}(k) = &-\frac{T}{2N} \sum_{q, l_1 l_2 l_3 l_4 } \Gamma^{\mathrm{s/t}}_{ l_1 l_3 l_4 l_2}(q) \\ &\times G_{l_3 l_5 } (k-q) G_{l_4 l_6 }(q-k) \Delta_{l_5 l_6 }(k-q),
\end{split}
   \label{gapeq}
\end{align}
where  $k \equiv (\bm{k},i\omega_n)$ and $q \equiv (\bm{q},i\nu_n)$ with $\bm{k}$ and $\bm{q}$ being the crystal momentum and $\omega_n$ ($\nu_n$) the fermionic (bosonic) Matsubara frequency. $T$ is temperature, $N$ the number of $\bm{q}$-points in the FBZ, $G(k)$ the bare Green's function, and $\Delta(k)$ the gap function. $\Gamma^{\mathrm{s/t}}_{ l_1 l_3 l_4 l_2}(q)$ describe the effective singlet (s) or triplet (t) pairing interaction accounting for scattering of electrons in orbitals $(l_1,l_2)$ with four-momenta $(k,-k)$ to $(l_3,l_4)$ with $(k-q,-k+q)$. In the case of singlet pairing, for instance, the pairing interaction reads \cite{bickers_self-consistent_2004}
\begin{align}  \label{gamma_s}
\mathbf{\Gamma}^{\mathrm{s}}(q) =  \frac{3\mathbf{\Gamma}^\mathrm{sp}}{2} +  \frac{\mathbf{\Gamma}^\mathrm{ch}}{2} + 3 \mathbf{\Gamma}^\mathrm{sp} \bm{\chi}^\mathrm{sp}(q) \mathbf{\Gamma}^\mathrm{sp} - \mathbf{\Gamma}^\mathrm{ch} \bm{\chi}^\mathrm{ch}(q) \mathbf{\Gamma}^\mathrm{ch},
\end{align}  
where $\bm{\chi}^{\mathrm{sp/ch}}$ are the spin (sp) and charge (ch) susceptibilities and $\mathbf{\Gamma}^{\mathrm{sp/ch}}$ the corresponding vertices (bold symbols will be used to denote matrices). $\bm{\chi}^{\mathrm{sp/ch}}(q)=\bm{\chi}^0(q) [ \mathbf{1} \mp \mathbf{\Gamma}^{\mathrm{sp/ch}} \bm{\chi}^0(q) ]^{-1}$ where $\mathbf{1}$ is the identity matrix and $\bm{\chi}^0(q)$ the irreducible susceptibility, whose elements are given by
\begin{align}  \label{chi0}
 \chi^0_{l_1 l_2 l_3 l_4}(q)	= -\frac{T}{N}\sum_{k}G_{l_1 l_3}(k+q)G_{l_4 l_2}(k).
\end{align}
$\mathbf{\Gamma}^{\mathrm{sp/ch}}$ are constructed from the onsite Kanamori Hamiltonian for the $e_g$ manifold (including intraorbital onsite Coulomb repulsion $U$, Hund’s coupling $J$, and the interorbital onsite Coulomb repulsion $U' = U - 2J$) and the interlayer density-density interaction  $V$ \cite{supple}. 
The phase transition to the superconducting state is indicated by the temperature $T_\mathrm{c}$ at which the maximum eigenvalue $\lambda_\mathrm{sc}$ reaches unity.  Thus, $\lambda_\mathrm{sc}$ serves as a proxy for the superconducting instability. We investigate the leading $\lambda_\mathrm{sc}$ at a fixed $T = 1/140~\mathrm{eV} \simeq 83$~K. To avoid divergence of the dominant spin susceptibility while ensuring a sizable $\lambda_\mathrm{sc}$, we choose an interaction strength of $U=0.96$~eV and set $J=U/6$ and $V=0$ for the main discussion. 
Note that we use the RPA here to assess strain and doping dependence of superconducting instabilities, their leading pairing symmetry, and the role of the fermiology, but not to quantitatively predict $T_\mathrm{c}$.
See Supplementary for more details~\cite{supple}.

\begin{figure*} [!htbp] 
	\includegraphics[width=1.0\textwidth, angle=0]{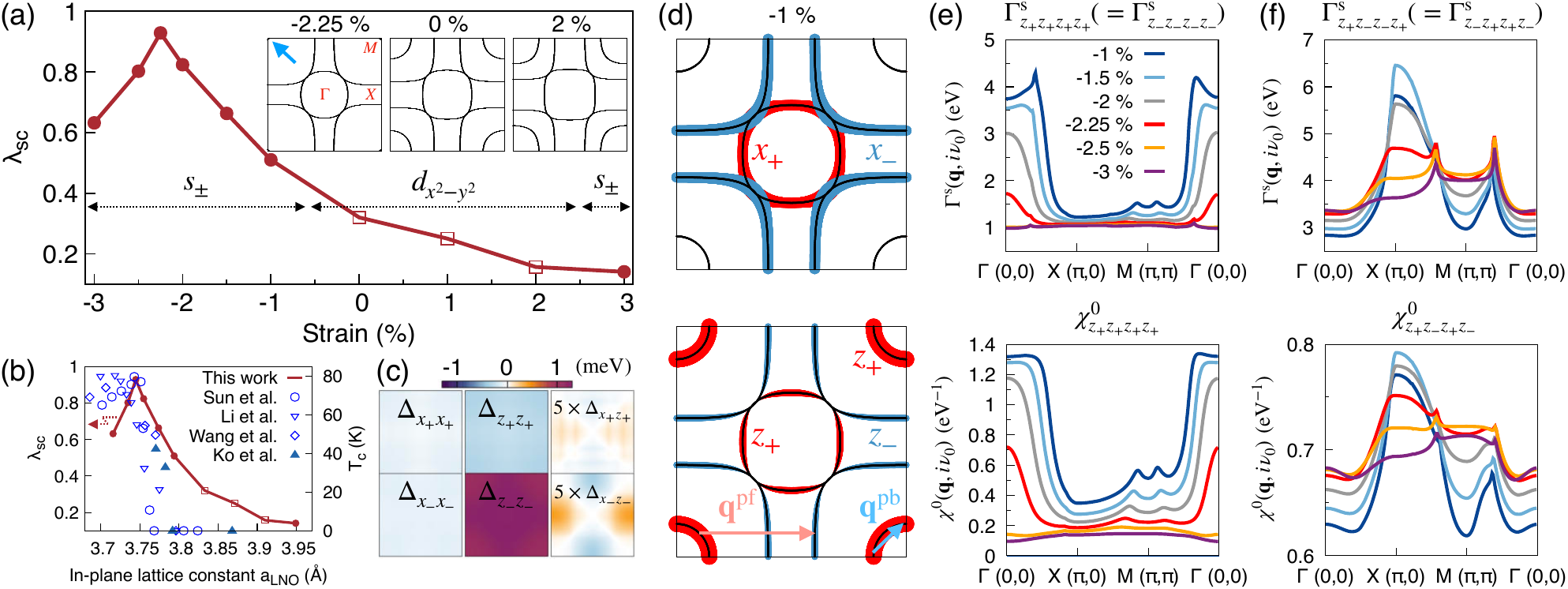}
	\caption{(a) The leading $\lambda_\mathrm{sc}$ as a function of strain. Circles: $s_\pm$-wave gap; squares: $d_{x^2-y^2}$-wave gap. The inset shows the FSs for several values of strain. The blue arrow highlights the marginal presence of the $\gamma$ pocket at $- 2.25$~\% strain. (b) Comparison between our calculated $\lambda_\mathrm{sc}$ (red line; left ticks) with experimental $T_\mathrm{c}$~\cite{sun_signatures_2023,li2025_100gpa,wang2024la2prni2o7,ko2025thinfilm} (right ticks) as a function of the in-plane lattice constant $a_\mathrm{LNO}$. Sun et al.: Ref.~\cite{sun_signatures_2023}; Li et al.: Ref.~\cite{li2025_100gpa}; Wang et al.: Ref.~\cite{wang2024la2prni2o7}; Ko et al.: Ref.~\cite{ko2025thinfilm}. The filled triangles denote the data for the strained thin films~\cite{ko2025thinfilm}, while the open symbols denote those for the pressurized samples~\cite{sun_signatures_2023,li2025_100gpa,wang2024la2prni2o7}.
    (c) The nonzero elements of the $s_\pm$-wave $\Delta_{lm}(\bm{k},i\omega_0)$ in the BA basis at $- 2.25$~\% strain. (d) Orbital weight on the FS at $- 1$~\% strain. The top panel shows the $x_+$ and $x_-$ contributions and the bottom panel shows the $z_+$ and $z_-$ contributions. 
    $\bm{q}^\mathrm{pb} = (\epsilon_1, \epsilon_2)$ with $\epsilon_1, \epsilon_2 < \pi$ being small numbers and $\bm{q}^\mathrm{pf} = (\pi, 0)$ denote wave vectors associated with the dominant pair-breaking and pair-forming channels, respectively, for the $s_\pm$-wave pairing. 
    (e) The pair-breaking interaction $\Gamma^\mathrm{s}_{z_+ z_+ z_+ z_+ }(\bm{q},i\nu_0) ~\{=\Gamma^\mathrm{s}_{z_- z_- z_- z_- }(\bm{q},i\nu_0) \}$ (top) and  the associated irreducible susceptibility $\chi^0_{z_+ z_+ z_+ z_+ }(\bm{q},i\nu_0)$ (bottom). (f) The pair-forming interaction $\Gamma^\mathrm{s}_{z_+ z_- z_- z_+ }(\bm{q},i\nu_0)~\{ =\Gamma^\mathrm{s}_{z_- z_+ z_+ z_- }(\bm{q},i\nu_0) \}$ (top) and the associated irreducible susceptibility $\chi^0_{z_+ z_- z_+ z_- }(\bm{q},i\nu_0)$ (bottom). 
	}
	\label{fig2}
\end{figure*}

Our main result is presented in Fig.~\ref{fig2}(a) which shows the leading $\lambda_\mathrm{sc}$ as a function of strain at an average electron filling of 3. We find that the instability to singlet pairing dominates over that to triplet pairing, and the maximum eigenvalue of $\bm{\chi}^{\mathrm{sp}}$ is much larger than that of $\bm{\chi}^{\mathrm{ch}}$ indicating that the singlet pairing is mediated by spin fluctuations. The dominant pairing symmetry for the compressive-strain side is $s_\pm$ where the gap changes sign between neighboring FS pockets \cite{supple}. This is consistent with the  experiment on the $-2$~\% strained sample, supporting the $s_\pm$ over $d$-wave gap symmetries~\cite{fan2025gapsymmetry}, and aligns with a recent theoretical study reporting the robustness of the $s_\pm$-wave pairing~\cite{ushio2025theoreticalstudyambientpressure}. 
While we find that the instability to the $d_{x^2-y^2}$-wave paring dominates over that to the $s_\pm$-wave under certain tensile strain, they are comparable in magnitude. This superconductivity–strain profile remains unaffected by variations in the interaction strength; see Supplementary for additional data~\cite{supple}.

Most notably, Fig.~\ref{fig2}(a) demonstrates that superconductivity is promoted by compressive strain, not by tensile strain, in agreement with experiments~\cite{ko2025thinfilm,zhou2025thinfilm,liu2025thinfilm,bhatt2025}.
Moreover, our calculated $\lambda_\mathrm{sc}$ exhibits a similar trend as that observed between $T_\mathrm{c}$ and the in-plane lattice constant $a_\mathrm{LNO}$ for both strained thin films~\cite{ko2025thinfilm} and pressurized samples~\cite{sun_signatures_2023,li2025_100gpa,wang2024la2prni2o7}, as shown in Fig.~\ref{fig2}(b). 
Note, however, that, as reported in Ref.~\cite{ko2025thinfilm}, the absolute value of $T_\mathrm{c}$ is also affected by the out-of-plane lattice constant which responds differently to compressive strain and hydrostatic pressure, so the agreement should not be overinterpreted.
In any case, these observations are surprising since the prevailing theoretical picture has been that the $\gamma$ pocket---which gets bigger with {\it tensile} strain (i.e., with increasing $a_\mathrm{LNO}$), as presented in the inset of Fig.~\ref{fig2}(a)---promotes spin-fluctuation-mediated superconductivity~\cite{sun_signatures_2023,zhang_trends_2023,lechermann_electronic_2023,yang_possible_2023,lu_superconductivity_2023,lu_interlayer_2023,tian_correlation_2023,yang_possible_2023,jiang2023high,fan2024superconductivity,zhang2024prediction,zhang2024spm,yang2024effective,zhang2024alternating,heier2024competing,xi2025_dwave,zhan2025impact,le2025landscape,yue2025correlated}.  Furthermore, $\lambda_{\mathrm{sc}}$ reaches its maximum at $-2.25$~\% strain, where the $\gamma$ pocket is marginally present at the $M$ point as shown in the FS in Fig.~\ref{fig2}(a).
 
To systematically trace the origin of this result, we first switch to the bonding-antibonding (BA) orbital basis. We hereafter denote the $d_{x^2-y^2}$ and $d_{z^2}$ orbitals by $\bar{x}$ and $\bar{z}$ for those in the top layer, and by $\underline{x}$ and $\underline{z}$ for those in the bottom layer, for brevity.
The BA orbitals ($x_\pm$ and $z_\pm$) are defined by the linear combinations of the top and bottom layer $e_g$ orbitals as 
\begin{align}
| x_{\pm} \rangle =  \frac{ | \bar{x} \rangle \pm  |  \underline{x} \rangle}{\sqrt{2}},\;\; | z_{\pm} \rangle =  \frac{ | \bar{z} \rangle \pm  |  \underline{z} \rangle}{\sqrt{2}},
\end{align}
where ket symbols indicate the corresponding MLWF orbitals in a unit cell.  A clear advantage of using this basis is that the BA orbitals for each type are orthogonal (i.e., $\langle x_+ | x_- \rangle=0$ and $\langle z_+ | z_- \rangle=0$),  and the $\gamma$ pocket consists entirely of the $z_+$ orbital character; see Fig.~\ref{fig2}(d).

A strikingly simple structure of the $s_\pm$-wave gap function emerges in the BA basis. As shown in Fig.~\ref{fig2}(c) for $-2.25$~\% strain, for instance, it exhibits significant magnitude only in the $\Delta_{z_+z_+}$ and $\Delta_{z_-z_-}$ components with opposite signs, indicating strong {\it interlayer} pairing between the top and bottom layer $z$ orbitals.
This observation allows us to simplify Eq.~(\ref{gapeq}) as
\begin{align} 
	\begin{split}
	\lambda_{\mathrm{sc}} \Delta_{z_+z_+}(k) \simeq &-T\Gamma^\mathrm{s}_{z_+ z_+ z_+ z_+}(q)|G_{z_+ z_+}(k')|^2 \Delta_{z_+z_+}(k') \\ &-T\Gamma^\mathrm{s}_{z_+ z_- z_- z_+}(q)|G_{z_- z_-}(k')|^2 \Delta_{z_- z_-}(k'), \\
		\lambda_{\mathrm{sc}} \Delta_{z_-z_-}(k) \simeq &-T\Gamma^\mathrm{s}_{z_- z_- z_- z_-}(q)|G_{z_- z_-}(k')|^2 \Delta_{z_-z_-}(k') \\ &-T\Gamma^\mathrm{s}_{z_- z_+ z_+ z_-}(q)|G_{z_+ z_+}(k')|^2 \Delta_{z_+ z_+}(k'),
	\end{split}    \label{gapeq2}
\end{align} 
with $k'\equiv k-q$, and the summation over $q$ and the associated prefactor $1/2N$ understood to be implicit on the right-hand sides of Eq~(\ref{gapeq2}). The pairing interactions appearing in Eq.~(\ref{gapeq2}) are all positive as shown in the top panels of Figs.~\ref{fig2}(e,~f) for the lowest bosonic Matsubara frequency $\nu_0=0$. 
Furthermore, exploiting the relation between $\mathbf{\Gamma}^\mathrm{s}$ [Eq.~(\ref{gamma_s})] and $\bm{\chi}^0$ [Eq.~(\ref{chi0})], and  noting that the spin susceptibility dominates over charge susceptibility, 
the relevant $\Gamma^\mathrm{s}$ components can be expanded up to second order in interactions as \cite{supple}
\begin{align}
\begin{split} \label{gamma_approx}
\Gamma^{\mathrm{s}}_{z_+ z_+ z_+ z_+}(q) &= \Gamma^{\mathrm{s}}_{z_- z_- z_- z_-}(q)  \simeq U + \frac{3}{4}U^2  \chi^0_{z_+ z_+ z_+ z_+}(q),   \\
\Gamma^{\mathrm{s}}_{z_+ z_- z_- z_+}(q) &= \Gamma^{\mathrm{s}}_{z_- z_+ z_+ z_-}(q)  \simeq U +  \frac{3}{2}U^2 \chi^0_{z_+ z_- z_+ z_-}(q),  
\end{split}
\end{align}
where terms containing $UJ$ and $J^2$ are dropped since $J \ll U$, and the term $ (3U^2/4)\chi^0_{z_- z_- z_- z_-}$ is neglected in the first line of Eq.~(\ref{gamma_approx}) because $\chi^0_{z_- z_- z_- z_-} \ll \chi^0_{z_+ z_+ z_+ z_+} $; see Supplementary for more details \cite{supple}.

The approximate gap equation [Eq.~(\ref{gapeq2})] along with Eq.~(\ref{gamma_approx}) 
provides key insight into two opposing pairing channels: To maximize $\lambda_{\mathrm{sc}}$, the first terms on the right-hand side of Eq.~(\ref{gapeq2}) should be minimized, while the second terms are maximized, since  $\Delta_{z_+z_+}<0$ and $\Delta_{z_-z_-}>0$ over the FBZ.  As a result, $\Gamma^{\mathrm{s}}_{z_+ z_+ z_+ z_+}$ and $\Gamma^{\mathrm{s}}_{z_- z_- z_- z_-}$ appearing in the first terms act as pair-breaking interactions, whereas $\Gamma^{\mathrm{s}}_{z_+ z_- z_- z_+}$ and  $\Gamma^{\mathrm{s}}_{z_- z_+ z_+ z_-}$ in the second terms serve as pair-forming interactions for the $s_\pm$-wave pairing. 
From Eq.~(\ref{gamma_approx}), we therefore identify $\chi^0_{z_+ z_+ z_+ z_+}$ and $\chi^0_{z_+ z_- z_+ z_-}$ as the main pair-breaking and pair-forming fluctuations, respectively. Figures~\ref{fig2}(e,~f) indeed show essentially the same momentum dependence of these $\chi^0$ elements as that of the associated pairing interactions. 
The corresponding momentum-transfer vectors clarify their microscopic origin~[Fig.~\ref{fig2}(d)]. Namely, the pair-breaking fluctuation $\chi^0_{z_+ z_+ z_+ z_+}$ is generated from intra-pocket scattering within the $\gamma$ pocket with $\bm{q}_\mathrm{pb} = (\epsilon_1 , \epsilon_2)$ with $\epsilon_{1,2} < \pi$, so it is active only when the $\gamma$ pocket crosses the Fermi level [Fig.~\ref{fig2}(e)]. In contrast, the pair-forming fluctuation $\chi^0_{z_+ z_- z_+ z_-}$ constitutes inter-pocket scattering involving not only the $\gamma$ but also the $\alpha$ and $\beta$ pockets, e.g., $\bm{q}_\mathrm{pf} = (\pi,0)$ between $\gamma$ and $\beta$, as shown in Fig.~\ref{fig2}(d). Due to its inter-pocket nature, it remains active even when the $\gamma$ pocket lies below the Fermi level [Fig.~\ref{fig2}(f)].

At this stage, we note that Eq.~(\ref{gapeq2}) reveals three key ingredients required to achieve a high $T_\mathrm{c}$ in general: (1) a strong pair-forming interaction, (2) a weak pair-breaking interaction, and (3) a large pair susceptibility $ \chi^\mathrm{p}_{\eta \eta}  \sim  T|G_{\eta \eta}|^2/2$ ($\eta \in \{z_+, z_-\}$), which scales the overall pairing strength. In our case, the dominant contribution arises from $\chi^\mathrm{p}_{z_+ z_+}$ since $|G_{z_+ z_+}|^2 \gg |G_{z_- z_-}|^2$; see Supplementary~\cite{supple}. We find that
\begin{align}
	\chi^\mathrm{p}_{z_+ z_+}(\bm{k}) = \frac{T}{2}\sum_{\omega_n}|G_{z_+ z_+}(\bm{k},i\omega_n)|^2 \sim \frac{\tanh \{ \xi_{\gamma \bm{k} } /(2T) \} }{2\xi_{\gamma \bm{k} }},
\end{align}
where $\xi_{\gamma \bm{k}}$ denotes the band energy of the $\gamma$ pocket with respect to the Fermi level. As the $\gamma$ pocket approaches and crosses the Fermi level from below, $\chi^\mathrm{p}_{z_+ z_+}$ increases, thereby amplifying the impact of both the pair-forming and pair-breaking channels that couple to it in Eq.~(\ref{gapeq2}).

Importantly, the resulting interplay between the two competing channels gives rise to a dichotomy across the Lifshitz transition. The pair-forming fluctuation $\chi^0_{z_+ z_- z_+ z_-}$  is weakly affected by the evolution of the $\gamma$ pocket and persists independently of whether this pocket crosses the Fermi level. 
In contrast, the pair-breaking fluctuation $\chi^0_{z_+ z_+ z_+ z_+}$ switches on sharply once the $\gamma$ pocket touches the Fermi level from below, and increases as this pocket rises further. It leads to the pronounced maximum of $\lambda_\mathrm{sc}$ at the Lifshitz transition point ($-2.25$~\% strain), as shown in Fig.~\ref{fig2}(a).
We note that our picture may have some relevance to the so-called ``incipient band'' scenario suggested in the context of iron-based superconductors or bilayer Hubbard models~\cite{kuroki2005,hirschfeld2011,chen2015_incipient,linscheid2016_incipient,Matsumoto2018_incipient}. Understanding of their connection requires further study.

A corollary of our Lifshitz scenario is that the presence of the $\gamma$ pocket on the FS is not a prerequisite for superconductivity in (La,Pr)$_3$Ni$_2$O$_7$. 
In fact, $\lambda_{\mathrm{sc}}$ exhibits an almost symmetric profile with respect to its peak at the Lifshitz transition point ($-2.25$~\% strain), where the $\gamma$ pocket touches the Fermi level. 
Therefore, superconductivity can emerge provided the ``valence-band maximum'' of the $\gamma$ pocket lies sufficiently close to the Fermi level, either from below or above. 
This finding not only reconciles two opposing viewpoints regarding which fermiology (i.e.,~three vs two pockets) underlies superconductivity~\cite{sun_signatures_2023,zhang_trends_2023,lechermann_electronic_2023,yang_possible_2023,lu_superconductivity_2023,lu_interlayer_2023,tian_correlation_2023,yang_possible_2023,jiang2023high,fan2024superconductivity,zhang2024prediction,zhang2024spm,yang2024effective,zhang2024alternating,heier2024competing,xi2025_dwave,zhan2025impact,le2025landscape,christiansson_correlated_2023,ryee2024_quenched,devaulx2025pressurestraineffectstextitab,yue2025correlated}, but also explains conflicting ARPES measurements that report the presence \cite{li2025arpes} and absence \cite{wang2025arpes} of the $\gamma$ pocket in compressively strained (La,Pr)$_3$Ni$_2$O$_7$ thin films exhibiting superconductivity.

On the tensile strain side, a sizable $\gamma$ pocket leads to an enhanced $\chi^0_{z_+ z_+ z_+ z_+}$ and thus $\Gamma^{\mathrm{s}}_{z_+ z_+ z_+ z_+}$ which significantly suppresses the $s_\pm$-wave pairing and instead promotes the $d_{x^2-y^2}$-wave pairing as the leading instability. 
However, due to the small $\lambda_{\mathrm{sc}}$ on the tensile-strain side [Fig.~\ref{fig2}(a)], an actual phase transition to superconductivity seems unlikely here.

\begin{figure} [!htbp] 
	\includegraphics[width=1.0\columnwidth, angle=0]{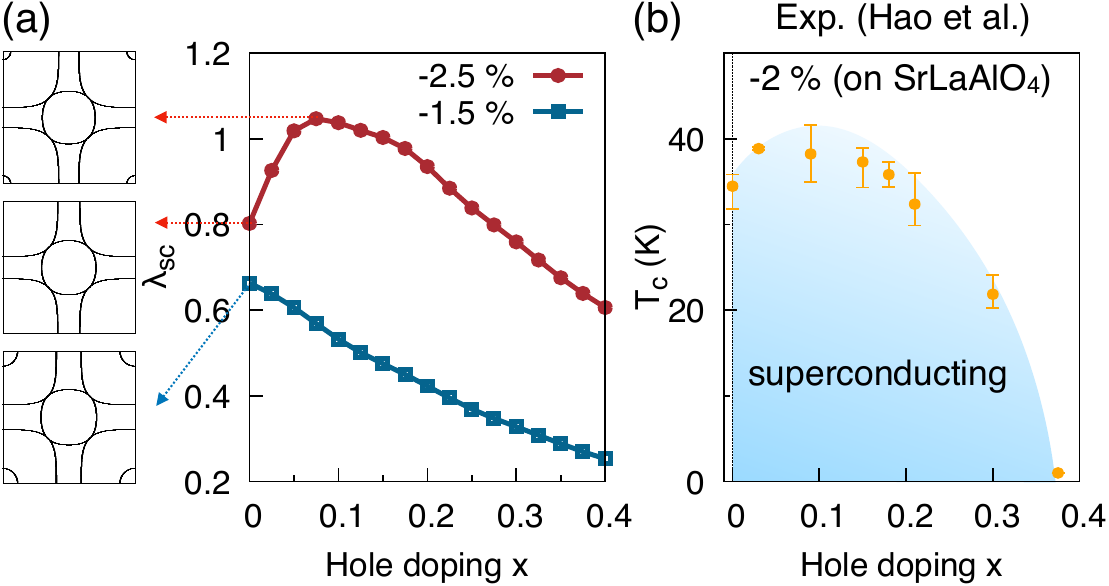}
	\caption{(a) $\lambda_\mathrm{sc}$ as a function of hole doping $x$ for $-2.5$~\% and $-1.5$~\% strain values (right), and the corresponding FSs for selected cases (left). (b) Experimental phase diagram of $-2$~\% strained La$_{3-x}$Sr$_{x}$Ni$_2$O$_7$ thin films grown on SrLaAlO$_4$ substrate with $x$ denoting hole doping. The yellow circles with error bars are the experimental data at which the resistivity reaches 98~\% of the extrapolated normal-state resistivity  \cite{hao2025superconductivityphasediagramsrdoped}.
	}
	\label{fig3}
\end{figure}

Having established the relationship between fermiology and superconducting instabilities as a function of strain, we now turn to the effect of doping. An interesting recent experimental observation \cite{hao2025superconductivityphasediagramsrdoped} is that $T_\mathrm{c}$ forms a dome shape as a function of hole doping $x$ in La$_{3-x}$Sr$_{x}$Ni$_2$O$_7$ thin films grown on SrLaAlO$_4$ substrate; see Fig.~\ref{fig3}(b). At an optimal hole doping of $x \simeq 0.09$, $T_\mathrm{c}$ reaches its maximum with $T_\mathrm{c} \simeq 42$~K \cite{hao2025superconductivityphasediagramsrdoped}. Motivated by this observation, we investigate the superconducting instability as a function of hole doping. To this end, we simply shift the chemical potential to achieve the desired doping level with the underlying MLWF band structure remaining unchanged at a given strain.

Figure~\ref{fig3}(a) shows the calculated $\lambda_{\mathrm{sc}}$  as a function of hole doping at two different compressive strain values. The leading instability remains also the $s_\pm$-wave pairing for the entire doping range.
Interestingly, two distinct doping dependencies emerge, depending on whether the $\gamma$ pocket is present at zero hole doping ($x=0$), or not. While $\lambda_{\mathrm{sc}}$ monotonically decreases with hole doping at $-1.5$~\% strain, it initially increases and then decreases with doping at $-2.5$~\% strain forming a maximum at $x = 0.075$.  This doping dependence in the $-2.5$~\% strain case is in agreement with the experiment \cite{hao2025superconductivityphasediagramsrdoped} shown in Fig.~\ref{fig3}(b).
Here as well, the maximum of $\lambda_{\mathrm{sc}}$ occurs when the $\gamma$ pocket is marginally present in the FS, as shown in the top left panel of Fig.~\ref{fig3}(a).  Enlarging the $\gamma$ pocket by further hole doping is detrimental to superconductivity, as it enhances the pair-breaking fluctuation $\chi^0_{z_+ z_+ z_+ z_+}$; see Supplementary for additional data and for data on the electron-doped region \cite{supple}.  
Thus, despite the simplified nature of our model, the proposed picture that $T_\mathrm{c}$ is optimal near a Lifshitz transition accounts for the overall trend of the experimental hole-doping dependence.

To conclude, we have established the Janus-faced role of the $\gamma$ pocket in spin-fluctuated-mediated superconductivity of strained bilayer nickelates. Our finding reconciles two seemingly contradicting perspectives on the fermiology (i.e., three vs two FS pockets), as well as accounts for several recent experimental observations in strained samples under ambient pressure. 
Very recently, an experiment on pressurized La$_3$Ni$_2$O$_7$ thin films reported superconductivity even under tensile strain at high pressures ($> \sim 10$~GPa)~\cite{osada2025}. Although the $\gamma$ pocket remains sizable under tensile strain~[Fig.~\ref{fig2}(a)], pressure can lead to a shrinking, or even quenching of the $\gamma$ pocket, assisted by interlayer electronic correlations~\cite{ryee2024_quenched}. We thus expect that the mechanism proposed in the present work can also be applied to this experimental finding.

A natural consequence of our picture is that optimal $T_\mathrm{c}$ is achieved when the $\gamma$ pocket surfaces at the Fermi level, placing the system close to a Lifshitz transition. This calls for future experimental investigations, and serves as a guiding principle for realizing higher $T_\mathrm{c}$. 
Establishing a unified framework of superconductivity that encompasses both strain and pressure dependence in the multilayer nickelate family remains an outstanding challenge, which requires more quantitative and sophisticated theoretical approaches.
The present work offers a conceptual basis for this endeavor.

{\it Acknowledgments.}~SR and TOW acknowledge funding and support by the DFG research unit FOR 5242 (WE 5342/7-1, project No. 449119662), and from the Cluster of Excellence ``CUI: Advanced Imaging of Matter'' of the DFG (EXC 2056, Project ID 390715994). NW and GS acknowledge funding and support by the DFG research unit FOR 5249 (``QUAST'', project No.~449872909) and the DFG funded SFB 1170 (``Tocotronics'', project No. 258499086). For this work the HPC-cluster Hummel-2 at University of Hamburg was used. The cluster was funded by Deutsche Forschungsgemeinschaft (DFG, German Research Foundation) – 498394658.

%\newpage
%\bibliographystyle{apsrev4-2}
%\bibliography{ref}


\end{document}

\renewcommand{\thepage}{S\arabic{page}}  
\renewcommand{\thesection}{SM\arabic{section}}   
\renewcommand{\thetable}{S\arabic{table}}   
\renewcommand{\thefigure}{S\arabic{figure}}
\renewcommand{\theequation}{S\arabic{equation}}

\renewcommand{\citenumfont}[1]{S#1}
\renewcommand{\bibnumfmt}[1]{[S#1]}

\def\tcb{\textcolor{blue}}
\def\tcr{\textcolor{red}}
\def\tcg{\textcolor{green}}
\def\tcc{\textcolor{cyan}}

\onecolumngrid

\title{Supplemental Material for \\ ``Superconductivity governed by Janus-faced fermiology in strained bilayer nickelates''}

\author{Siheon Ryee}
 \email{siheonryee@gmail.com} 
\affiliation{I. Institute of Theoretical Physics, University of Hamburg, Notkestra{\ss}e 9-11, 22607 Hamburg, Germany}
\affiliation{The Hamburg Centre for Ultrafast Imaging, Luruper Chaussee 149, 22761 Hamburg, Germany}

\author{Niklas Witt}
\affiliation{Institut f\"ur Theoretische Physik und Astrophysik and W\"urzburg-Dresden Cluster of Excellence ct.qmat, Universit\"at W\"urzburg, 97074 W\"urzburg, Germany}

\author{Giorgio Sangiovanni}
\affiliation{Institut f\"ur Theoretische Physik und Astrophysik and W\"urzburg-Dresden Cluster of Excellence ct.qmat, Universit\"at W\"urzburg, 97074 W\"urzburg, Germany}

\author{Tim O. Wehling}
\affiliation{I. Institute of Theoretical Physics, University of Hamburg, Notkestra{\ss}e 9-11, 22607 Hamburg, Germany}
\affiliation{The Hamburg Centre for Ultrafast Imaging, Luruper Chaussee 149, 22761 Hamburg, Germany}

\maketitle
\tableofcontents
\hypersetup{linkcolor=red}

\newpage
%\twocolumngrid

\section{Model construction for strained bilayer nickelates} \label{SM1}

We first performed DFT calculations using the Vienna Ab initio Simulation Package (VASP)~\cite{Kresse1993, Kresse1996, Kresse1996b} using DFT-optimized lattice constants of strained La$_3$Ni$_2$O$_7$ reported in Ref.~\cite{zhao2025_strain}. 
In Ref.~\cite{zhao2025_strain}, the in-plane ($a$ and $b$) and out-of-plane ($c$) lattice constants of the experimentally relevant Fmmm structure \cite{bhatt2025} are reported for strain of $-3$~\% and $3$~\%. Noting that the DFT-optimized lattice constants of the Amam structure vary nearly linearly with strain \cite{zhao2025_strain}, we performed a linear interpolation of the Fmmm lattice constants between $-3$~\% and $3$~\% to estimate the values at intermediate strain values. Since the in-plane lattice constants $a$ and $b$ only differ by about $1~\%$ in the orthorhombic Fmmm structure~\cite{zhao2025_strain,bhatt2025}, we use higher-symmetry tetragonal I4/mmm structure by averaging $a$ and $b$. 
Atomic positions were relaxed until the force was smaller than $1$~meV/\AA. 
We used $7 \times 7 \times 7$ and $15 \times 15 \times 15$ momentum grids for the atomic relaxation and electronic structure calculations, respectively, for the primitive unit cell at each strain.
We employed the Perdew–Burke–Ernzerhof (PBE) parametrization of the generalized gradient approximation for the exchange-correlation functional \cite{PBE} as in Ref.~\cite{zhao2025_strain} and the plane-wave energy cutoff of $500$~eV. 

We note that employing DFT+$U$ (instead of DFT), as done in many previous studies on bilayer nickelates, can lead to a different critical strain at which the $\gamma$ pocket disappears. With an appropriate choice of Coulomb interaction parameters and double-counting scheme, this approach can potentially yield better quantitative agreement with experimental data. However, since the aim of our study is not to quantitatively track the evolution of the FS, but rather to gain a qualitative understanding of the relationship between fermiology and superconducting instabilities, the parameter-free DFT should be sufficient for our purpose.

\begin{table} [!htbp] 
\caption{Hopping amplitudes in units of eV.} \label{hk_table}
\renewcommand{\arraystretch}{1.6}
\begin{centering}
\setlength{\tabcolsep}{7pt}  
\begin{tabular}{c|r|r|r|r|r|r|r|r|r|r}  
\hline
\hline
\multicolumn{1}{c|}{$\quad \quad$} & 
\multicolumn{10}{c}{Strain} \\
%\hline
\multicolumn{1}{c|}{} &
\multicolumn{1}{c}{$-3$~\%} & 
\multicolumn{1}{c}{$-2.5$~\%} & 
\multicolumn{1}{c}{$-2.25$~\%} & 
\multicolumn{1}{c}{$-2$~\%} & 
\multicolumn{1}{c}{$-1.5$~\%} & 
\multicolumn{1}{c}{$-1$~\%} & 
\multicolumn{1}{c}{0~\%} & 
\multicolumn{1}{c}{1~\%} & 
\multicolumn{1}{c}{2~\%} & 
\multicolumn{1}{c}{3~\%} \\
\hline
$\varepsilon_0$  &   0.7306   &  0.6467    &  0.6045     &  0.5556    &  0.4963    &  0.4308    &  0.3049    & 0.1891     & 0.0856     &   -0.0105           \\
$t^x_1$  &  -0.4999    &  -0.4925    &  -0.4884    &  -0.4829    &  -0.4756    & -0.4666     &  -0.4486    & -0.4313    &  -0.4148    &     -0.3986         \\
$t^x_2$  &   0.0727   &  0.0714    & 0.0708     & 0.0700     &  0.0690    &  0.0677    & 0.0656     &  0.0637    &  0.0617 &     0.0594         \\
$t^x_3$  &  -0.0544  &  -0.0524    & -0.0514     &  -0.0501    & -0.0489     & -0.0477     & -0.0455     &  -0.0430    & -0.0405     &    -0.0385          \\
$t^z_1$  &  -0.0825    & -0.0870     & -0.0892 &  -0.0914    &   -0.0947    & -0.0998     & -0.1085     & -0.1162 &  -0.1232    &     -0.1306         \\
$t^z_2$  &  -0.0142    &  -0.0141    &  -0.0142    &  -0.0143    &  -0.0148    &  -0.0158    &  -0.0174    &  -0.0184    & -0.0194     &     -0.0206         \\
$t^{x}_{\perp}$  &  -0.0021    & -0.0023     &  -0.0023    & -0.0025     &   -0.0029   & -0.0036     & -0.0052     &  -0.0062    &  -0.0073    &       -0.0085      \\
$t^{x}_{\perp,1}$  & -0.0036     & -0.0035     &  -0.0033    &  -0.0031    &  -0.0031    &  -0.0032    & -0.0031     & -0.0030   & -0.0030     &    -0.0031          \\
$t^{x}_{\perp,2}$  &  0.0018     & 0.0019     &  0.0019    &  0.0020    &  0.0021    &  0.0023    & 0.0027     &  0.0030    &  0.0031    &     0.0033         \\
$t^{z}_{\perp}$  &  -0.6408    &  -0.6424    & -0.6418     &  -0.6406    & -0.6380     &  -0.6265    & -0.6048     &  -0.5909    &  -0.5832    &    -0.5752          \\
$t^{z}_{\perp,1}$  & 0.0049     & 0.0074     & 0.0087     &  0.0102    & 0.0126     & 0.0165     & 0.0235     &  0.0283    &  0.0316    &    0.0343          \\
$t^{z}_{\perp,2}$  & 0.0080     & 0.0086     &  0.0086    & 0.0088     &  0.0088    &  0.0084    & 0.0077     &  0.0072    &  0.0073    &    0.0072          \\
$t^{xz}_1$  &  0.2084    &  0.2103    &  0.2113    & 0.2124     & 0.2145     & 0.2178     & 0.2227     & 0.2253     &  0.2273    &    0.2292          \\
$t^{xz}_3$  &  0.0242    &  0.0235    &  0.0232    & 0.0228     & 0.0226     & 0.0226     &  0.0224    &0.0221    &  0.0216    &      0.0213        \\
$t^{xz}_{\perp,1}$  &  -0.0276    & -0.0286     &  -0.0291    &  -0.0291    & -0.0295     & -0.0302     &  -0.0307    & -0.0309     &  -0.0311    &     -0.0313         \\
\hline
\hline
\end{tabular}
\end{centering} 
\end{table}

We then used the Wannier90 code~\cite{Wanner90} to construct a four-orbital (top-layer two Ni-$e_g$ + bottom-layer two Ni-$e_g$) bilayer square-lattice model of strained La$_3$Ni$_2$O$_7$ using the maximally localized Wannier function (MLWF) method \cite{marzari2012}. We hereafter denote the $d_{x^2-y^2}$ orbital by $x$ and the $d_{z^2}$ orbital by $z$ for brevity. The resulting four-orbital tight-binding Hamiltonian $H_0$ on the bilayer square lattice reads
\begin{align}
H_0 = \sum_{\bm{k}\sigma}\Psi^\dagger_{\bm{k}\sigma}\bm{h}(\bm{k}) \Psi_{\bm{k}\sigma},
\end{align}
where $\Psi_{\bm{k}\sigma} = [d_{\bm{k}\bar{x}\sigma},d_{\bm{k}\bar{z}\sigma},d_{\bm{k}\underline{x}\sigma},d_{\bm{k}\underline{z}\sigma}]^T$ with $\bar{\eta}$ and $\underline{\eta}$ ($\eta \in \{ x,z \}$) referring to the orbital in the top and bottom layers, respectively. $d$ denotes the corresponding electron annihilation operator. $\sigma$ denotes spin and $\bm{k}$ the crystal momentum in the first Brillouin zone. $\bm{h}(\bm{k})$ reads
\begin{align}
\bm{h}(\bm{k}) =
\left(
\renewcommand{\arraystretch}{1.5} 
\begin{array}{cccc}
h_{11}(\bm{k}) & h_{12}(\bm{k}) & h_{13}(\bm{k}) & h_{14}(\bm{k}) \\
h_{12}(\bm{k}) & h_{22}(\bm{k}) & h_{23}(\bm{k}) & h_{24}(\bm{k}) \\
h_{13}(\bm{k}) & h_{23}(\bm{k}) & h_{33}(\bm{k}) & h_{34}(\bm{k}) \\
h_{14}(\bm{k}) & h_{24}(\bm{k}) & h_{34}(\bm{k}) & h_{44}(\bm{k})
\end{array}
\right)
\end{align}
where the matrix elements of $\bm{h}(\bm{k})$ are given by
\begin{align} \label{hk}
\begin{split}
h_{11}(\bm{k}) &= h_{33}(\bm{k}) = \frac{\varepsilon_0}{2} + 2t^x_{1} (\cos k_x + \cos k_y)  + 4t^{x}_{2} \cos k_x \cos k_y + 2t^{x}_{3} (\cos 2k_x + \cos 2k_y),  \\
h_{22}(\bm{k}) &= h_{44}(\bm{k}) = -\frac{\varepsilon_0}{2}  + 2t^z_{1} (\cos k_x + \cos k_y)  + 4t^{z}_{2} \cos k_x \cos k_y , \\
h_{12}(\bm{k}) &= 2t^{xz}_{1} (\cos k_x - \cos k_y) + 2t^{xz}_{3} (\cos 2k_x - \cos 2k_y),  \\
h_{13}(\bm{k}) &= t^{x}_{\perp} + 2t^{x}_{\perp,1} (\cos k_x + \cos k_y) + 4t^{x}_{\perp,2} \cos k_x \cos k_y ,  \\
h_{14}(\bm{k}) &= 2t^{xz}_{\perp,1} (\cos k_x - \cos k_y),  \\
h_{24}(\bm{k}) &= t^{z}_{\perp} + 2t^{z}_{\perp,1} (\cos k_x + \cos k_y) + 4t^{z}_{\perp,2} \cos k_x \cos k_y ,  \\
h_{23}(\bm{k}) &= h_{14}(\bm{k}),  \\
h_{34}(\bm{k}) &= h_{12}(\bm{k}).
\end{split}
\end{align}
The hopping amplitudes are listed in Table~\ref{hk_table}.

\section{Pairing interactions and the gap equation}

\subsection{Pairing interactions}

We consider the following interaction Hamiltonian $H_\mathrm{int}$ for the bilayer $e_g$ manifold:
\begin{align} \label{kanamori}
\begin{split} 
	H_\mathrm{int} &= U \sum_{\eta} ( n_{\bar{\eta}\uparrow} n_{\bar{\eta}\downarrow} + n_{\underline{\eta}\uparrow} n_{\underline{\eta}\downarrow} )  + \sum_{\substack{\eta < \eta' \\ \sigma, \sigma'}} (U' - J \delta_{\sigma\sigma'}) 
(n_{\bar{\eta}\sigma} n_{\bar{\eta}'\sigma'} + n_{\underline{\eta}\sigma} n_{\underline{\eta}'\sigma'}) \\
&\quad + J \sum_{\eta \ne \eta'} \left( 
d^{\dagger}_{\bar{\eta}\uparrow} d^{\dagger}_{\bar{\eta}'\downarrow} d_{\bar{\eta}\downarrow} d_{\bar{\eta}'\uparrow} 
+ d^{\dagger}_{\bar{\eta}\uparrow} d^{\dagger}_{\bar{\eta}\downarrow} d_{\bar{\eta}'\downarrow} d_{\bar{\eta}'\uparrow} 
+ d^{\dagger}_{\underline{\eta}\uparrow} d^{\dagger}_{\underline{\eta}'\downarrow} d_{\underline{\eta}\downarrow} d_{\underline{\eta}'\uparrow} 
+ d^{\dagger}_{\underline{\eta}\uparrow} d^{\dagger}_{\underline{\eta}\downarrow} d_{\underline{\eta}'\downarrow} d_{\underline{\eta}'\uparrow} 
\right) \\
&\quad + V \sum_{\substack{\eta, \eta' \\ \sigma, \sigma'}} n_{\bar{\eta}\sigma} n_{\underline{\eta}'\sigma'}  
\end{split}
\end{align}
where $\eta,\eta' \in \{ x, z  \}$ and $\sigma,\sigma' \in \{ \uparrow, \downarrow \}$. $\bar{\eta}$ and $\underline{\eta}$ denote the orbital in the top and bottom layers, respectively. $n_{\eta\sigma}=d^\dagger_{\eta\sigma} d_{\eta\sigma}$ is the corresponding electron number operator.
The first three terms constitute the Kanamori-type local interaction Hamiltonian which includes intraorbital onsite Coulomb repulsion $U$, Hund’s coupling $J$, and the interorbital onsite Coulomb repulsion $U' = U - 2J$. The last term, including the interlayer density–density interaction $V$, accounts for nonlocal interlayer interaction between the $e_g$ orbitals in the top and bottom layers within the same cell. Equation~(\ref{kanamori}) can be recast as
\begin{align} \label{kanamori2}
\begin{split} 
	H_\mathrm{int} &= \frac{1}{2}\sum_{i, l_1 l_2 l_3 l_4, \sigma \sigma'} \mathcal{U}^{\sigma \sigma'}_{l_1 l_3 l_4 l_2} d^{\dagger}_{i l_1 \sigma} d^{\dagger}_{i l_2 \sigma'} d_{i l_4 \sigma'} d_{i l_3 \sigma}\;,
\end{split}
\end{align}
where $l_i \in \{\bar{x}, \bar{z}, \underline{x}, \underline{z} \}$. Here $U \equiv \mathcal{U}^{\sigma \sigma}_{\bar{\eta} \bar{\eta} \bar{\eta} \bar{\eta}} = \mathcal{U}^{\sigma -\sigma}_{\bar{\eta} \bar{\eta} \bar{\eta} \bar{\eta}} =\mathcal{U}^{\sigma \sigma}_{\underline{\eta} \underline{\eta} \underline{\eta} \underline{\eta}} =\mathcal{U}^{\sigma -\sigma}_{\underline{\eta} \underline{\eta} \underline{\eta} \underline{\eta}} $, $U' \equiv \mathcal{U}^{\sigma \sigma}_{\bar{\eta} \bar{\eta}  \bar{\eta}' \bar{\eta}' } = \mathcal{U}^{\sigma -\sigma}_{\bar{\eta} \bar{\eta} \bar{\eta}' \bar{\eta}' } = \mathcal{U}^{\sigma \sigma}_{\underline{\eta} \underline{\eta} \underline{\eta}'  \underline{\eta}' } = \mathcal{U}^{\sigma -\sigma}_{\underline{\eta} \underline{\eta} \underline{\eta}'  \underline{\eta}' } $, and $J  \equiv  \mathcal{U}^{\sigma \sigma}_{\bar{\eta} \bar{\eta}' \bar{\eta}' \bar{\eta}} =  \mathcal{U}^{\sigma -\sigma}_{\bar{\eta} \bar{\eta}' \bar{\eta}' \bar{\eta}} = \mathcal{U}^{\sigma \sigma}_{\bar{\eta} \bar{\eta}' \bar{\eta} \bar{\eta}'}=\mathcal{U}^{\sigma -\sigma}_{\bar{\eta} \bar{\eta}' \bar{\eta} \bar{\eta}'} =  \mathcal{U}^{\sigma \sigma}_{\underline{\eta} \underline{\eta}' \underline{\eta}' \underline{\eta}} = \mathcal{U}^{\sigma -\sigma}_{\underline{\eta} \underline{\eta}' \underline{\eta}' \underline{\eta}}=  \mathcal{U}^{\sigma \sigma}_{\underline{\eta} \underline{\eta}' \underline{\eta} \underline{\eta}'} = \mathcal{U}^{\sigma -\sigma}_{\underline{\eta} \underline{\eta}' \underline{\eta} \underline{\eta}'}$ with $\eta \neq \eta'$. $V \equiv  \mathcal{U}^{\sigma \sigma}_{\bar{\eta} \bar{\eta}  \underline{\eta}' \underline{\eta}' } = \mathcal{U}^{\sigma -\sigma}_{\bar{\eta} \bar{\eta} \underline{\eta}' \underline{\eta}' } = \mathcal{U}^{\sigma \sigma}_{\underline{\eta} \underline{\eta} \bar{\eta}'  \bar{\eta}' } = \mathcal{U}^{\sigma -\sigma}_{\underline{\eta} \underline{\eta} \bar{\eta}'  \bar{\eta}' }$ where either $\eta \neq \eta'$ or $\eta = \eta'$.

From Eq.~(\ref{kanamori2}), elements of the spin (sp) and charge (ch) channel vertices $\bm{\Gamma}^{\mathrm{sp/ch}}$ within the random phase approximation (RPA) are given by \cite{bickers_self-consistent_2004}
\begin{align}
	\Gamma^{\mathrm{sp/ch}}_{l_1 l_2 l_3 l_4}  = \mathcal{U}_{l_1 l_2 l_3 l_4}^{\uparrow \downarrow} \mp (\mathcal{U}_{l_1 l_2 l_3 l_4}^{\uparrow \uparrow} - \mathcal{U}_{l_1 l_3 l_2 l_4}^{\uparrow \uparrow}),
\end{align}
where the terms in parentheses ensure the anti-symmetrization of the Coulomb vertex. In the $e_g$-orbital basis, elements of  $\bm{\Gamma}^{\mathrm{sp/ch}}$  read
\begin{align}
	\Gamma^{\mathrm{sp}}_{\bar{\eta}_1 \bar{\eta}_2 \bar{\eta}_3 \bar{\eta}_4} 
	= \Gamma^{\mathrm{sp}}_{\underline{\eta}_1 \underline{\eta}_2 \underline{\eta}_3 \underline{\eta}_4} 
	= \left\{
	\begin{aligned}
		&U  \\ 
		&U' \\ 
		&J  \\ 
		&J
	\end{aligned}
	\right.\;,\quad    
	\Gamma^{\mathrm{ch}}_{\bar{\eta}_1 \bar{\eta}_2 \bar{\eta}_3 \bar{\eta}_4} 
	= \Gamma^{\mathrm{ch}}_{\underline{\eta}_1 \underline{\eta}_2 \underline{\eta}_3 \underline{\eta}_4} 
	= \left\{
	\begin{aligned}
		&U &&\mathrm{if}\, \eta_1= \eta_2 = \eta_3 = \eta_4,\\
		&-U' + 2J &&\mathrm{if}\, \eta_1 = \eta_3 \neq \eta_2 = \eta_4, \\
		&2U' -J &&\mathrm{if}\, \eta_1 = \eta_2 \neq \eta_3 = \eta_4, \\
		&J &&\mathrm{if}\, \eta_1 = \eta_4 \neq \eta_2 = \eta_3,
	\end{aligned}
	\right.
\end{align}
and 
\begin{align}
	\Gamma^{\mathrm{sp}}_{\bar{\eta} \underline{\eta}' \bar{\eta} \underline{\eta}'} 
	= \Gamma^{\mathrm{sp}}_{\underline{\eta} \bar{\eta}' \underline{\eta} \bar{\eta}'} 
	= V~\;,\quad    
	\Gamma^{\mathrm{ch}}_{\bar{\eta} \underline{\eta}' \bar{\eta} \underline{\eta}'} 
	= \Gamma^{\mathrm{ch}}_{\underline{\eta} \bar{\eta}' \underline{\eta} \bar{\eta}'} 
	= -V\;,\quad   
    \Gamma^{\mathrm{ch}}_{\bar{\eta} \bar{\eta} \underline{\eta}'  \underline{\eta}'} 
    = \Gamma^{\mathrm{ch}}_{\underline{\eta} \underline{\eta} \bar{\eta}'  \bar{\eta}'} 
    = 2V\;,
\end{align}
where either $\eta = \eta'$ or $\eta \neq \eta'$. The other remaining elements are zero. 
The spin and charge susceptibilities are obtained using these vertices from the Bethe-Salpeter equation as
\begin{align}
	\chi^{\mathrm{sp/ch}}_{l_1  l_2 l_7  l_8}(q)  &= \chi^{0}_{l_1  l_2 l_7  l_8}(q) \pm  \chi^{\mathrm{sp/ch}}_{l_1  l_2 l_3  l_4}(q) \Gamma^{\mathrm{sp/ch}}_{l_3 l_4 l_5 l_6} \chi^{0}_{l_5  l_6 l_7  l_8}(q), \\ &= \big[ \bm{\chi}^{0}(q)[\bm{1} \mp \bm{\Gamma}^{\mathrm{sp/ch}} \bm{\chi}^{0}(q)]^{-1} \big]_{l_1 l_2 l_7 l_8},
	\label{chi}
\end{align} 
where $\chi^{0}_{l_1 l_2 l_3 l_4}(q) = -\frac{T}{N}\sum_{\bm{k}}G_{l_2 l_3}(k+q)G_{l_4 l_2}(k)$. $k \equiv (\bm{k},i\omega_n)$ and $q \equiv (\bm{q},i\nu_n)$ with $\bm{k}$ and $\bm{q}$ being the crystal momentum and $\omega_n$ ($\nu_n$) the fermionic (bosonic) Matsubara frequency. $T$ is temperature, $N$ the number of $\bm{k}$ points in the first Brillouin zone, and $G(k)$ the Green's function. Here, indices repeated twice should be summed over.
By employing the parquet equations \cite{bickers_self-consistent_2004,rohringer_diagrammatic_2018} and using $\bm{\Gamma}^{\mathrm{sp/ch}}$ and $\bm{\chi}^{\mathrm{sp/ch}}$, one can express the effective RPA singlet (s) and triplet (t) pairing interactions appearing in Eq.~(1) in the main text as 
\begin{align}
\begin{split}
\Gamma^{\mathrm{s}}_{l_1 l_3 l_4 l_2}(q)
&= \frac{3}{2}{\Gamma}^\mathrm{sp}_{l_1 l_3 l_4 l_2} + \frac{1}{2}{\Gamma}^\mathrm{ch}_{l_1 l_3 l_4 l_2}  + 3[\bm{\Gamma}^\mathrm{sp} \bm{\chi}^\mathrm{sp}(q) \bm{\Gamma}^\mathrm{sp}]_{l_1 l_3 l_4 l_2} - [\bm{\Gamma}^\mathrm{ch} \bm{\chi}^\mathrm{ch}(q) \bm{\Gamma}^\mathrm{ch}]_{l_1 l_3 l_4 l_2},
\end{split}\label{Gamma_s}  \\
\begin{split}
\Gamma^{\mathrm{t}}_{l_1 l_3 l_4 l_2}(q)
&= -\frac{1}{2}{\Gamma}^\mathrm{sp}_{l_1 l_3 l_4 l_2} + \frac{1}{2}{\Gamma}^\mathrm{ch}_{l_1 l_3 l_4 l_2}  - [\bm{\Gamma}^\mathrm{sp} \bm{\chi}^\mathrm{sp}(q) \bm{\Gamma}^\mathrm{sp}]_{l_1 l_3 l_4 l_2} - [\bm{\Gamma}^\mathrm{ch} \bm{\chi}^\mathrm{ch}(q) \bm{\Gamma}^\mathrm{ch}]_{l_1 l_3 l_4 l_2}. 
\end{split}
\end{align}

\subsection{Solving the gap equation}
The transition to a superconducting phase is indicated by the divergence of the corresponding pairing susceptibility.
Using the above formulas, it can be formulated in terms of a non-Hermitian eigenvalue problem in the normal state, namely the gap equation [Eq.~(1) in the main text]~\cite{bickers_self-consistent_2004}, which reads
\begin{align} \label{gapeq}
	\lambda_{\mathrm{sc}} \Delta_{l_1 l_2}(k) = &-\frac{T}{2N} \sum_{q, l_1 l_2 l_3 l_4 } \Gamma^{\mathrm{s/t}}_{ l_1 l_3 l_4 l_2}(q) G_{l_3 l_5 } (k-q) G_{l_4 l_6 }(q-k) \Delta_{l_5 l_6 }(k-q),
\end{align} 
where $\lambda_\mathrm{sc}$ is the eigenvalue and $\Delta_{l_1 l_2}(k)$ the gap function. We investigate the leading $\lambda_\mathrm{sc}$ at a fixed $T = 1/140~\mathrm{eV} \simeq 83$~K using $140 \times 140$ momentum ($\bm{k}$ and $\bm{q}$) grids in the first Brillouin zone.

\begin{figure} [!htbp] 
	\includegraphics[width=0.5\textwidth, angle=0]{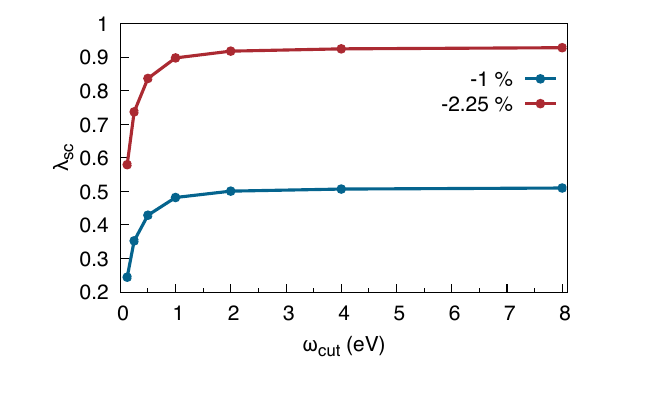}
	\caption{The calculated leading eigenvalue $\lambda_\mathrm{sc}$ at $-1$~\% and  $-2.25$~\% strain values as a function of the cutoff energy $\omega_\mathrm{cut}$ for the fermionic Matsubara frequency in $\Delta(\bm{k}, i\omega)$. $U=0.96$~eV, $J=U/6$ and $V=0$.
	}
	\label{sfig_lambda_conv}
\end{figure}

We employed the implicitly restarted Arnoldi method using the ARPACK library~\cite{lehoucq1998arpack} to compute the leading eigenvalue and the associated gap function of Eq.~(\ref{gapeq}). We set a cutoff energy $\omega_\mathrm{cut}=8$~eV for the fermionic Matsubara frequency in $\Delta(\bm{k}, i\omega)$ to ensure that the matrix size in the gap equation remains numerically tractable without losing accuracy. As a result, $\Delta(\bm{k}, i\omega)$ is defined only within the frequency range of $[-\omega_\mathrm{cut},\omega_\mathrm{cut}]$. $\lambda_\mathrm{sc}$ is found to be nearly converged at $\omega_\mathrm{cut} \simeq 2$~eV (see Fig.~\ref{sfig_lambda_conv}), suggesting that $\Delta(\bm{k}, i\omega)$ has significant magnitude only within the Matsubara frequency range of approximately $[-2, 2]$~eV and decays rapidly with increasing $|i\omega|$.

The gap functions must satisfy the following transformation properties under the exchange of spin ($\hat{S}$), parity ($\hat{P}$), orbital ($\hat{O}$), and time ($\hat{T}$)~\cite{Linder2019}: 
\begin{align}
\hat{S} \Delta_{l_1 l_2}(\bm{k}, i\omega) &= \pm \Delta_{l_1 l_2}(\bm{k}, i\omega), 
\quad \text{(+ for triplet, $-$ for singlet)} \\
\hat{P} \Delta_{l_1 l_2}(\bm{k}, i\omega) &= \pm \Delta_{l_1 l_2}(-\bm{k}, i\omega), \\
\hat{O} \Delta_{l_1 l_2}(\bm{k}, i\omega) &= \pm \Delta_{l_2 l_1}(\bm{k}, i\omega), \\
\hat{T} \Delta_{l_1 l_2}(\bm{k}, i\omega) &= \pm \Delta_{l_1 l_2}(\bm{k}, -i\omega).
\end{align}
The anti-symmetry of the gap function, dictated by the Pauli principle, leads to the constraint:
\begin{align}
\hat{S}\hat{P}\hat{O}\hat{T} \Delta_{l_1 l_2}(\bm{k}, i\omega) = -\Delta_{l_1 l_2}(\bm{k}, i\omega). 
\end{align}
We explored all possible symmetries consistent with the above “SPOT” condition when solving Eq.~(\ref{gapeq}). Note that the $s_\pm$-wave and $d_{x^2-y^2}$-wave gap functions are both odd under $\hat{S}$ (i.e.,~singlet), and even under $\hat{P}$, $\hat{O}$, and $\hat{T}$, while they belong to different irreducible representations ($A_{1g}$ for the $s_\pm$ wave and $B_{1g}$ for the $d_{x^2-y^2}$ wave).

\section{Derivation of Eq.~(6) in the main text}

\subsection{Basis transformation}
The bonding-antibonding (BA) orbital basis in which the electron annhilation operator $d_{\eta_\pm}$ (site and spin indices are omitted for simplicity) is defined as the symmetric ($+$) and antisymmetric ($-$) combinations of the top and bottom layer $e_g$ orbital operators:
\begin{align}
\begin{pmatrix}
	d_{x_+ } \\
	d_{z_+ } \\	
	d_{x_- } \\
	d_{z_- } 
\end{pmatrix} 
= \frac{1}{\sqrt{2}} \begin{pmatrix}  1 & 0 & 1 & 0 \\ 0 & 1 & 0 & 1 \\  1 & 0 & -1 & 0 \\ 0 & 1 & 0 & -1 \end{pmatrix} 
\begin{pmatrix}
	d_{\bar{x} } \\
	d_{\bar{z} } \\	
	d_{\underline{x} } \\
	d_{\underline{z} } 
\end{pmatrix}
= \bm{A} 
\begin{pmatrix}
	d_{\bar{x} } \\
	d_{\bar{z} } \\	
	d_{\underline{x} } \\
	d_{\underline{z} } 
\end{pmatrix}, 
~\mathrm{where}~\bm{A} \equiv \frac{1}{\sqrt{2}} \begin{pmatrix}  1 & 0 & 1 & 0 \\ 0 & 1 & 0 & 1 \\  1 & 0 & -1 & 0 \\ 0 & 1 & 0 & -1 \end{pmatrix}.
\end{align}
Note that $\bm{A}^{-1} = \bm{A}$. Under this basis transformation, the vertices $\bm{\Gamma}^r$ (with $r = \mathrm{s}, \mathrm{t}, \mathrm{sp}, \mathrm{ch}$) and susceptibilities $\bm{\chi}^r$ (with $r = 0, \mathrm{sp}, \mathrm{ch}$) are transformed into the BA basis as
\begin{align} \label{G_ba}
\Gamma^r_{ijkl}(q) = \sum_{abcd}A_{ia}A_{jb}A_{kc}A_{ld}\Gamma^r_{abcd}(q),\quad \chi^r_{ijkl}(q) = \sum_{abcd}A_{ia}A_{jb}A_{kc}A_{ld}\chi^r_{abcd}(q),
\end{align}
where $i,j,k,l \in \{ x_+, z_+, x_-, z_-  \}$ and $a,b,c,d \in \{ \bar{x}, \bar{z}, \underline{x}, \underline{z}  \}$.

\subsection{Eq.~(6) in the main text}
\begin{figure} [!htbp] 
	\includegraphics[width=0.8\textwidth, angle=0]{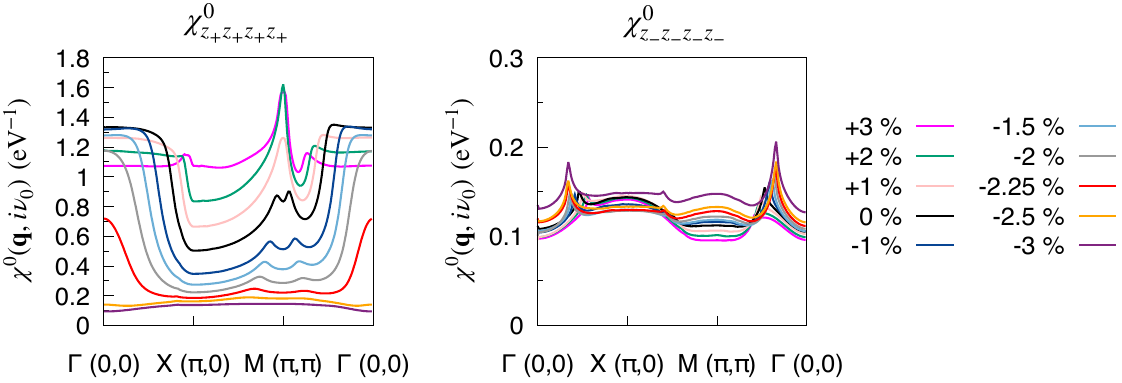}
	\caption{The irreducible susceptibilities $\chi^0_{z_+ z_+ z_+ z_+ }(\bm{q}, i\nu_0)$ (left) and  $\chi^0_{z_- z_- z_- z_- }(\bm{q}, i\nu_0)$ (right), at the lowest bosonic Matsubara frequency $\nu_0 = 0$, presented over the entire strain range.
	}
	\label{sfig_chi0}
\end{figure}

We now investigate which irreducible susceptibility governs the relevant singlet pairing interactions.
The charge susceptibility $\bm{\chi}^\mathrm{ch}$ is found to be much smaller than the spin susceptibility $\bm{\chi}^\mathrm{sp}$, so terms containing $\bm{\chi}^\mathrm{ch}$ can be neglected for our purpose. 
Expanding the spin susceptibility $\bm{\chi}^\mathrm{sp}$ in terms of the irreducible (bubble) susceptibility $\bm{\chi}^0(q)$, we can express Eq.~(\ref{Gamma_s}) as
\begin{align}
\begin{split} \label{eq.Gs}
    \bm{\Gamma}^{\mathrm{s}}(q) &\simeq \frac{3}{2}\bm{\Gamma}^\mathrm{sp} + \frac{1}{2}\bm{\Gamma}^\mathrm{ch}  + 3 \bm{\Gamma}^\mathrm{sp} \bm{\chi}^\mathrm{sp}(q) \bm{\Gamma}^\mathrm{sp} 
    \\ &= \frac{3}{2}\bm{\Gamma}^\mathrm{sp} + \frac{1}{2}\bm{\Gamma}^\mathrm{ch} + 3 \bm{\Gamma}^\mathrm{sp} \big[ \bm{\chi}^0(q) + \bm{\chi}^0(q) \bm{\Gamma}^\mathrm{sp} \bm{\chi}^0(q) +  \bm{\chi}^0(q) \bm{\Gamma}^\mathrm{sp} \bm{\chi}^0(q) \bm{\Gamma}^\mathrm{sp} \bm{\chi}^0(q) +...  \big] \bm{\Gamma}^\mathrm{sp}.
\end{split}
\end{align}
Retaining terms up to second order in interactions, this expression reduces to
\begin{align}
\begin{split} \label{eq.Gs2}
    \bm{\Gamma}^{\mathrm{s}}(q) &\simeq \frac{3}{2}\bm{\Gamma}^\mathrm{sp} + \frac{1}{2}\bm{\Gamma}^\mathrm{ch} + 3 \bm{\Gamma}^\mathrm{sp} \bm{\chi}^0(q) \bm{\Gamma}^\mathrm{sp}.
\end{split}
\end{align}
We are interested in $\Gamma^{\mathrm{s}}_{z_+ z_+ z_+ z_+}$, $\Gamma^{\mathrm{s}}_{z_- z_- z_- z_-}$, $\Gamma^{\mathrm{s}}_{z_+ z_- z_- z_+}$, and $\Gamma^{\mathrm{s}}_{z_- z_+ z_+ z_-}$ appearing in Eq.~(5) in the main text. Using Eqs.~(\ref{G_ba}) and (\ref{eq.Gs2}), and neglecting second order terms containing $J$ since $J\ll U$, we can express them as
\begin{align}
\begin{split} \label{eq.Gs3}
\Gamma^{\mathrm{s}}_{z_+ z_+ z_+ z_+}(q) &\simeq U + \frac{3}{4}V + \frac{3}{4}(U+V)^2 \chi^0_{z_+ z_+ z_+ z_+}(q) + \frac{3}{4}(U-V)^2\chi^0_{z_- z_- z_- z_-}(q) \\ &\simeq U + \frac{3}{4}V + \frac{3}{4}(U+V)^2 \chi^0_{z_+ z_+ z_+ z_+}(q),  \\
\Gamma^{\mathrm{s}}_{z_- z_- z_- z_-}(q) &\simeq U + \frac{3}{4}V + \frac{3}{4}(U+V)^2 \chi^0_{z_- z_- z_- z_-}(q) + \frac{3}{4}(U-V)^2\chi^0_{z_+ z_+ z_+ z_+}(q)\\ &\simeq U + \frac{3}{4}V + \frac{3}{4}(U-V)^2\chi^0_{z_+ z_+ z_+ z_+}(q),   \\
\Gamma^{\mathrm{s}}_{z_+ z_- z_- z_+}(q) &= \Gamma^{\mathrm{s}}_{z_- z_+ z_+ z_-}(q) \simeq U + \frac{3}{4}V + \frac{3}{4}(U^2-V^2) \big[\chi^0_{z_+ z_- z_+ z_-}(q) + \chi^0_{z_- z_+ z_- z_+}(q)\big] \\ &= U + \frac{3}{4}V + \frac{3}{2}(U^2-V^2) \chi^0_{z_+ z_- z_+ z_-}(q),
\end{split}
\end{align}
where we have omitted the term containing $\chi^0_{z_- z_- z_- z_-}(q)$ since $\chi^0_{z_- z_- z_- z_-}(q) \ll \chi^0_{z_+ z_+ z_+ z_+}(q)$. Although it is strictly valid for strain values $>-2.5$~\% [cf.~Fig.~\ref{sfig_chi0}], neglecting $\chi^0_{z_- z_- z_- z_-}(q)$ over the entire strain range does not affect our discussion. We used the identity $\chi^0_{z_+ z_- z_+ z_-}(q) = \chi^0_{z_- z_+ z_- z_+}(q)$.
In the case where $V=0$, the above Eq.~(\ref{eq.Gs3}) reduces to 
\begin{align}
\begin{split}
\Gamma^{\mathrm{s}}_{z_+ z_+ z_+ z_+}(q) &= \Gamma^{\mathrm{s}}_{z_- z_- z_- z_-}(q) \simeq U + \frac{3}{4}U^2  \chi^0_{z_+ z_+ z_+ z_+}(q),  \\
\Gamma^{\mathrm{s}}_{z_+ z_- z_- z_+}(q) &= \Gamma^{\mathrm{s}}_{z_- z_+ z_+ z_-}(q) \simeq U +  \frac{3}{2}U^2 \chi^0_{z_+ z_- z_+ z_-}(q),
\end{split}
\end{align}
which corresponds to Eq.~(6) in the main text.

\section{Influence of Hund's coupling and interlayer interactions}
\begin{figure} [!htbp] 
	\includegraphics[width=0.85\textwidth, angle=0]{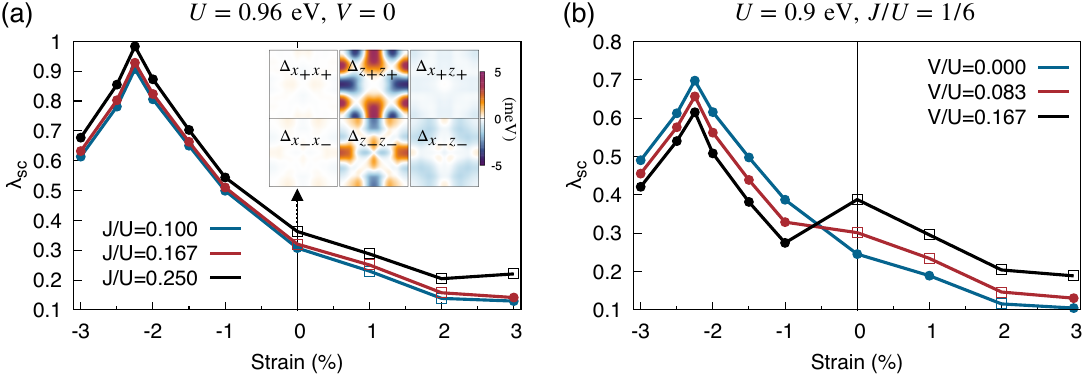}
	\caption{(a) The leading $\lambda_\mathrm{sc}$ as a function of strain (bottom) for different values of $J/U$ at $U=0.96$~eV and $V=0$. The inset shows the nonzero elements of the $d_{x^2-y^2}$-wave gap function $\Delta_{lm}(\bm{k},i\omega_0)$ in the BA basis at $0$~\% strain. (b) The leading $\lambda_\mathrm{sc}$ as a function of strain (bottom) for different values of $V/U$ at $U=0.9$~eV and $J=U/6 \simeq 0.167U$. Circles: $s_\pm$-wave gap; squares: $d_{x^2-y^2}$-wave gap.
	}
	\label{sfig_lambda}
\end{figure}

Figure~\ref{sfig_lambda} illustrates the dependence of the leading superconducting eigenvalue $\lambda_\mathrm{sc}$ on (a) Hund’s coupling $J$ and (b) the interlayer density-density interaction $V$. 

The influence of $J$ is found to be minor within a physically reasonable range. On the tensile strain side, increasing $J$ promotes $d_{x^2-y^2}$-wave pairing as the leading instability. In contrast, on the compressive strain side, the leading pairing instability is always the $s_\pm$-wave gap.

The effect of $V$ differs between the compressive and tensile strain regimes: while $\lambda_\mathrm{sc}$ decreases with increasing $V$ under compressive strain, it tends to increase at zero strain and under tensile strain. This behavior arises because $V$ enhances $\Gamma^\mathrm{s}_{z_+ z_+ z_+ z_+}$, which acts as a pair-breaking interaction for the $s_\pm$-wave pairing (compressive strain side), but as a pair-forming interaction for the $d_{x^2 - y^2}$-wave pairing (tensile strain side). This can be understood from recalling Eq.~(\ref{eq.Gs3}) which reads 
\begin{align}
\Gamma^\mathrm{s}_{z_+ z_+ z_+ z_+}(q) \simeq U + \frac{3}{4}V + \frac{3}{4}(U+V)^2 \chi^0_{z_+ z_+ z_+ z_+}(q).
\end{align}
Thus, $\Gamma^\mathrm{s}_{z_+ z_+ z_+ z_+}(\bm{q})$ increases with $V$, and the contribution from the irreducible susceptibility $\chi^0_{z_+ z_+ z_+ z_+}(\bm{q})$ becomes more significant as $V$ grows.
As shown in Fig.~\ref{sfig_chi0}(a), $\chi^0_{z_+ z_+ z_+ z_+}(\bm{q})$ develops a peak at the $M$ point [$(\pi, \pi)$] of the FBZ on the tensile strain sides. For the $d_{x^2 - y^2}$-wave pairing symmetry [see the inset of Fig.~\ref{sfig_lambda}(a)], this momentum $\bm{q} = (\pi, \pi)$ connects two regions of the dominant gap component $\Delta_{z_+ z_+}$ that have opposite signs. Consequently, the associated fluctuation $\chi^0_{z_+ z_+ z_+ z_+}$ and the resulting $\Gamma^\mathrm{s}_{z_+ z_+ z_+ z_+}$ acts as a pair-forming channel for the $d_{x^2 - y^2}$-wave pairing. 

This behavior---namely, the suppression of $s_\pm$-wave pairing and the enhancement of $d_{x^2 - y^2}$-wave pairing with increasing $V$ is consistent with a recent study using the fluctuation-exchange approximation~\cite{xi2025_dwave}.

\section{Why $\lambda_{\mathrm{sc}}$ is maximized near a Lifshitz transition in the BCS framework}
\begin{figure} [!htbp] 
		\includegraphics[width=1.0\textwidth, angle=0]{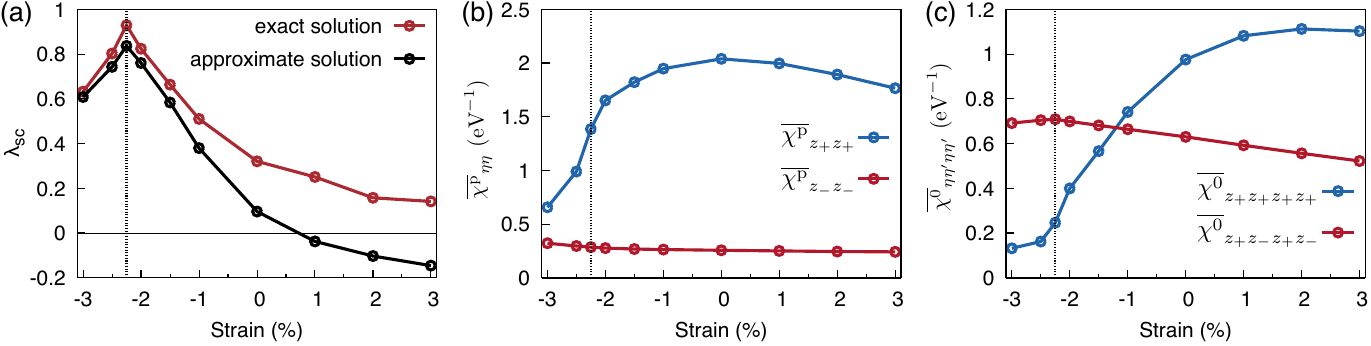}
		\caption{(a) The calculated $\lambda_\mathrm{sc}$ as a function of strain, obtained by solving the full gap equation [Eq.~(1) in the main text] (red line) and from the approximate analytical expression [Eq.~(\ref{lambda_approx})]. (b) Momentum-averaged pair susceptibilities. (c) Momentum-averaged pair-breaking fluctuation ($\overline{\chi^0}_{z_+ z_+ z_+ z_+}$) and pair-forming fluctuation ($\overline{\chi^0}_{z_+ z_- z_+ z_-}$). 
			The vertical dotted lines in (a--c) denote the strain ($-2.25$~\%) at which $\lambda_\mathrm{sc}$ is maximal. $U=0.96$~eV, $J=U/6$, and $V=0$. 
		}
		\label{sfig_approx}
	\end{figure}

To further understand why $\lambda_{\mathrm{sc}}$ is maximized near a Lifshitz transition, we start from the approximate gap equation in the main text [Eq.~(5)], which reads:
\begin{align} 
	\begin{split}
		\lambda_{\mathrm{sc}} \Delta_{z_+z_+}(k) \simeq & \frac{T}{2N}\sum_{q} \Big[ -\Gamma^\mathrm{s}_{z_+ z_+ z_+ z_+}(q)|G_{z_+ z_+}(k-q)|^2 \Delta_{z_+z_+}(k-q) -\Gamma^\mathrm{s}_{z_+ z_- z_- z_+}(q)|G_{z_- z_-}(k-q)|^2 \Delta_{z_- z_-}(k-q) \Big], \\
		\lambda_{\mathrm{sc}} \Delta_{z_-z_-}(k) \simeq & \frac{T}{2N}\sum_{q} \Big[ -\Gamma^\mathrm{s}_{z_- z_- z_- z_-}(q)|G_{z_- z_-}(k-q)|^2 \Delta_{z_-z_-}(k-q) -\Gamma^\mathrm{s}_{z_- z_+ z_+ z_-}(q)|G_{z_+ z_+}(k-q)|^2 \Delta_{z_+ z_+}(k-q) \Big]. \label{gap_approx0}
	\end{split} 
\end{align}
Our aim is to derive a simple BCS-like equation starting from this equation. To this end, we assume that the $s$-wave gap functions $ \Delta_{z_+z_+}(k)$ and $ \Delta_{z_- z_-}(k)$ are both momentum- and frequency-independent, namely $ \Delta_{z_+z_+}(k) \to \Delta_{z_+z_+}$ and $\Delta_{z_- z_-}(k) \to \Delta_{z_- z_-}$.  Furthermore, as in the usual BCS case, we assume the pairing interactions ($\Gamma^\mathrm{s}$) are frequency-independent (i.e., $\Gamma^\mathrm{s}(\bm{q},i\nu_n) = \Gamma^\mathrm{s}(\bm{q},0)$ for all $n$). We then take the average over $\bm{k}$ (i.e., $\frac{1}{N}\sum_{\bm{k}}$) on both sides of the above equation. It leads to
\begin{align} 
	\begin{split}
		\lambda^\mathrm{approx}_{\mathrm{sc}} \Delta_{z_+z_+} & =  - V^\mathrm{pb}_{z_+ z_+}  \Big( \frac{T}{2N} \sum_{\bm{k}', \omega_n} |G_{z_+ z_+}(\bm{k}',i\omega_n)|^2 \Big) \Delta_{z_+z_+}  -  V^\mathrm{pf}_{z_+ z_-}  \Big( \frac{T}{2N} \sum_{\bm{k}', \omega_n} |G_{z_- z_-}(\bm{k}',i\omega_n)|^2 \Big) \Delta_{z_- z_-},    \\
		\lambda^\mathrm{approx}_{\mathrm{sc}} \Delta_{z_-z_-} & = - V^\mathrm{pb}_{z_- z_-}  \Big( \frac{T}{2N} \sum_{\bm{k}', \omega_n} |G_{z_- z_-}(\bm{k}',i\omega_n)|^2 \Big) \Delta_{z_- z_-}  -  V^\mathrm{pf}_{z_- z_+}  \Big( \frac{T}{2N} \sum_{\bm{k}', \omega_n} |G_{z_+ z_+}(\bm{k}',i\omega_n)|^2 \Big) \Delta_{z_+ z_+}, \label{gap_approx1}
	\end{split} 
\end{align}
where we have replaced $\simeq$ with $=$ in going from Eq.~(\ref{gap_approx0}) to Eq.~(\ref{gap_approx1}) and accordingly $\lambda_{\mathrm{sc}}$ with $\lambda^\mathrm{approx}_{\mathrm{sc}}$. $\bm{k}' \equiv \bm{k} - \bm{q}$.  We have also defined momentum-averaged pair-breaking (pb) and pair-forming (pf) interactions in Eq.~(\ref{gap_approx1}) as
\begin{align}  
	%	\begin{split}
		V^\mathrm{pb}_{z_+ z_+} &\equiv \frac{1}{N}\sum_{\bm{q}} \Gamma^\mathrm{s}_{z_+ z_+ z_+ z_+}(\bm{q},0),\;\;	V^\mathrm{pb}_{z_- z_-} \equiv \frac{1}{N}\sum_{\bm{q}} \Gamma^\mathrm{s}_{z_- z_- z_- z_-}(\bm{q},0), \label{Vpb} \\
		V^\mathrm{pf}_{z_+ z_-} &\equiv  \frac{1}{N}\sum_{\bm{q}} \Gamma^\mathrm{s}_{z_+ z_- z_+ z_-}(\bm{q},0),\;\; V^\mathrm{pf}_{z_- z_+} \equiv \frac{1}{N}\sum_{\bm{q}} \Gamma^\mathrm{s}_{z_- z_+ z_- z_+}(\bm{q},0). \label{Vpf}
		%\end{split}
	\end{align}
	The terms in parentheses in Eq.~(\ref{gap_approx1}) can be reduced to the usual BCS kernel after performing the summation over the Matsubara frequency $\omega_n$ as follows:
	\begin{align}  
		\overline{\chi^\mathrm{p}}_{\eta \eta} &\equiv \frac{T}{2N} \sum_{\bm{k}', \omega_n} |G_{\eta \eta}(\bm{k}',i\omega_n)|^2 \\ 
		&\simeq  \frac{T}{2N} \sum_{l, \bm{k}', \omega_n} \frac{1}{\omega_n^2 + \xi_{l\bm{k}' }^2} |\langle \phi_{\eta \bm{k}' } | \psi_{l \bm{k}' } \rangle|^4 \\ 
		&= \frac{1}{2N} \sum_{l, \bm{k}'} \frac{\tanh \big( \xi_{l\bm{k}' } /(2T) \big)}{2\xi_{l\bm{k}' }}  |\langle \phi_{\eta \bm{k}' } | \psi_{l \bm{k}' } \rangle|^4 \label{kernel}, 
	\end{align}
	where $\eta \in \{ z_+, z_-\}$. $\xi_{l\bm{k}'}$ denotes the band energy of band $l$ at momentum $\bm{k}'$, and $|\psi_{l\bm{k}'}\rangle$ the corresponding eigenstate.
	The quantity $|\langle \phi_{\eta\bm{k}'} | \psi_{l\bm{k}'} \rangle|^2$ represents the weight of the orbital $\eta$ in the eigenstate $|\psi_{l\bm{k}'}\rangle$, obtained by projecting it onto the $\eta$-orbital Wannier state $|\phi_{\eta\bm{k}'}\rangle$. Equation~(\ref{kernel}) takes the form of the familiar BCS kernel or the momentum-averaged pair susceptibility; hence, we denoted it as $\overline{\chi^\mathrm{p}}_{\eta \eta}$. When evaluating Eq.~(\ref{kernel}), $E_{l\bm{k}'}$ is assumed to be finite within an energy window $[-\Lambda, \Lambda]$ as in the usual BCS theory. %where this cutoff is determined by a Debye frequency. 
	For demonstration purposes, we set $\Lambda = 2$~eV here. Now, Eq.~(\ref{gap_approx1}) can be written in the form of the BCS-like gap equation as
	\begin{align} 
		\begin{split}
			\lambda^\mathrm{approx}_{\mathrm{sc}} \Delta_{z_+z_+} = & - \underbrace{V^\mathrm{pb}_{z_+ z_+}  \overline{\chi^\mathrm{p}}_{z_+ z_+}}_{=\lambda^\mathrm{pb}_{z_+ z_+}} \Delta_{z_+z_+}  -  \underbrace{V^\mathrm{pf}_{z_+ z_-}  \overline{\chi^\mathrm{p}}_{z_- z_-}}_{=\lambda^\mathrm{pf}_{z_+ z_-}}  \Delta_{z_- z_-} , \\
			\lambda^\mathrm{approx}_{\mathrm{sc}} \Delta_{z_-z_-} = & - \underbrace{V^\mathrm{pb}_{z_- z_-}  \overline{\chi^\mathrm{p}}_{z_- z_-}}_{=\lambda^\mathrm{pb}_{z_- z_-}} \Delta_{z_- z_-}  - \underbrace{V^\mathrm{pf}_{z_- z_+}  \overline{\chi^\mathrm{p}}_{z_+ z_+}}_{=\lambda^\mathrm{pf}_{z_- z_+}}  \Delta_{z_+z_+}. \label{gap_approx2}
		\end{split} 
	\end{align}
	Solving the above coupled equation [Eq.~(\ref{gap_approx2})], we obtain
	\begin{align}
		\lambda^\mathrm{approx}_\mathrm{sc} = \frac{ -\big( \lambda^\mathrm{pb}_{z_+ z_+} + \lambda^\mathrm{pb}_{z_- z_-} \big) + \sqrt{ \big( \lambda^\mathrm{pb}_{z_+ z_+} + \lambda^\mathrm{pb}_{z_- z_-} \big)^2 +  4 \big( \lambda^\mathrm{pf}_{z_+ z_-} \lambda^\mathrm{pf}_{z_- z_+}  - \lambda^\mathrm{pb}_{z_+ z_+} \lambda^\mathrm{pb}_{z_- z_-} \big) }}{2}.  \label{lambda_approx}
	\end{align}
	Note that $\lambda^\mathrm{pb}_{z_+ z_+}$ and $\lambda^\mathrm{pb}_{z_- z_-}$ represent the effective pair-breaking strengths, whereas $\lambda^\mathrm{pf}_{z_+ z_-}$ and $\lambda^\mathrm{pf}_{z_- z_+}$ the effective pair-forming strengths.  These effective strengths are dimensionless.
	$\lambda^\mathrm{approx}_\mathrm{sc}$ is plotted as a function of strain with a black line in Fig.~\ref{sfig_approx}(a), together with the exact solution obtained by solving Eq.~(1) in the main text. We find that $\lambda^\mathrm{approx}_\mathrm{sc}$ obtained from Eq.~(\ref{lambda_approx}) perfectly follows the trend of the exact solution, although its absolute value deviates due to the simplified nature of Eq.~(\ref{gap_approx2}). %---namely, neglecting the $x_\pm$ orbitals and averaging over the momentum space.  
	Below, we investigate Eq.~(\ref{gap_approx2}) to gain a qualitative understanding of the physics underlying the exact solution.

	Now we address the question of ``Why is $\lambda_{\mathrm{sc}}$ maximized near a Lifshitz transition?''.
Note first that the pair susceptibility $\overline{\chi^\mathrm{p}}_{z_+ z_+}$ dominates over $\overline{\chi^\mathrm{p}}_{z_- z_-}$, reflecting the contribution from the $z_+$-character $\gamma$ pocket which has a high density of states due to its flat-band nature near its valence-band maximum.
In addition, $\overline{\chi^\mathrm{p}}_{z_+ z_+}$ monotonically increases from $-3$~\% to $0$~\% strain [Fig.~\ref{sfig_approx}(b)]~\footnote{The enhancement of $\overline{\chi^\mathrm{p}}_{z_+ z_+}$ as the $\gamma$ pocket approaches the Fermi level from below can be understood in analogy to the BCS case by assuming a constant density of states $D_0$  and using $|\langle \phi_{z_+ \bm{k}' } | \psi_{\gamma \bm{k}' } \rangle|^2=1$ in Eq.~(\ref{kernel}). It converts the summation over $\bm{k}'$ to an energy integral:
	\begin{align}
		\overline{\chi^\mathrm{p}}_{z_+ z_+} \simeq \frac{D_0}{2} \int_{-\Lambda}^{\xi^\mathrm{VBM}_\gamma} {\mathrm{d}\xi_\gamma} \frac{\tanh \big( \xi_\gamma/(2T) \big)}{2\xi_\gamma} \overset{|\xi^\mathrm{VBM}_\gamma| \gg T}{\simeq} \frac{D_0}{4} \ln{\frac{\Lambda}{|\xi^\mathrm{VBM}_\gamma|}} \label{log},
	\end{align}
	where $\xi^\mathrm{VBM}_\gamma$ refers to the valence-band maximum of the $\gamma$ pocket, with $\xi^\mathrm{VBM}_\gamma < 0$. Thus, it shows that $\overline{\chi^\mathrm{p}}_{z_+ z_+}$ increases logarithmically as the $\gamma$ pocket approaches the Fermi level from below as in Fig.~\ref{sfig_approx}(b), even when $|\xi^\mathrm{VBM}_\gamma|$ far exceeds the thermal energy.}, thereby amplifying the impact of both the pair-forming and pair-breaking channels that couple to it in Eq.~(\ref{gap_approx2}).

The key lies in the dichotomy between the pair-breaking and pair-forming fluctuations, $\chi^0_{z_+z_+z_+z_+}$ and $\chi^0_{z_+ z_- z_+ z_-}$, respectively. The pair-breaking fluctuation $\chi^0_{z_+ z_+ z_+ z_+}$ involves only the $\gamma$ pocket and is therefore active only when this pocket crosses the Fermi level. In contrast, the pair-forming fluctuation $\chi^0_{z_+ z_- z_+ z_-}$ constitutes inter-pocket scattering involving not only the $\gamma$, but also the $\alpha$ and $\beta$ pockets, e.g., $\bm{q}_\mathrm{pf} = (\pi,0)$ between $\gamma$ and $\beta$, as shown in Fig.~2(d) in the main text. Due to its inter-pocket nature, it remains active even when the $\gamma$ pocket lies below the Fermi level.

This contrasting behavior of the two distinct fluctuation channels explains why $\lambda_\mathrm{sc}$ (or $\lambda^\mathrm{approx}_\mathrm{sc}$) reaches its maximum at the Lifshitz transition.
Namely, as the $\gamma$ pocket passes through the Fermi level from below, the pair-forming interaction itself changes weakly, but its effect is amplified by the enhanced pair susceptibility $\chi^\mathrm{p}_{z_+ z_+}$, enhancing $\lambda_\mathrm{sc}$ (or $\lambda^\mathrm{approx}_\mathrm{sc}$) . However, the pair-breaking interaction switches on abruptly when the $\gamma$ pocket crosses the Fermi level at $-2.25$~\% strain and increases as this pocket rises further, causing $\lambda_\mathrm{sc}$ (or $\lambda^\mathrm{approx}_\mathrm{sc}$)  to decrease beyond the Lifshitz transition point.

\section{Additional data: Gap function on the Fermi surface}
\begin{figure} [!htbp] 
	\includegraphics[width=0.6\textwidth, angle=0]{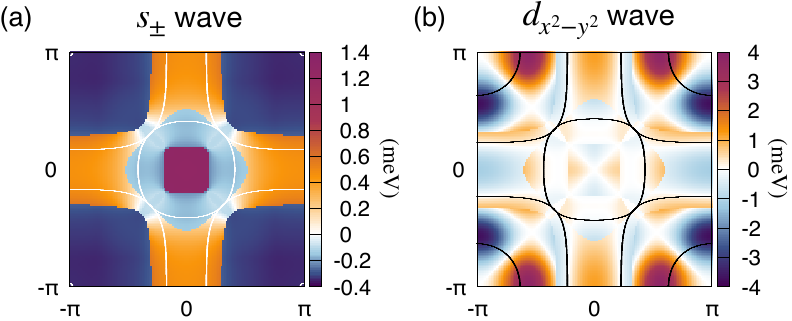}
	\caption{(a) The $s_\pm$-wave gap at $U=0.96$~eV, $J=U/6$, and $V=0$ for $-2.25$~\% strain. (a) The $d_{x^2-y^2}$-wave gap at $U=0.9$~eV and $J=V=U/6$ for $0$~\% strain. The white solid lines in (a) and black solid lines in (b) indicate the corresponding FSs.}
	\label{sfig_FS}
\end{figure}

Figure~\ref{sfig_FS} shows the gap function on the band $n$ closest to the Fermi level, $\hat{\Delta}_n(\bm{k}, i\omega_0)$ ($\omega_0 = \pi T$ is the lowest fermionic Matsubara frequency). It is obtained by projecting the gap function $\Delta_{l_1 l_2}(\bm{k}, i\omega_0)$ in the $e_g$-basis onto the band $n$ closest to the Fermi level at each momentum $\bm{k}$: $\hat{\Delta}_n(\bm{k},i\omega_0)=\sum_{l_1 l_2}P^{-1}_{n l_1}\Delta_{l_1l_2}(\bm{k},i\omega_0)P_{l_2 n}$ where $P^{-1}_{n l_1}$ and $P_{l_2 n}$ are elements of the basis transformation matrices from the $e_g$ orbitals $l_1$ and $l_2$ to the band $n$, respectively.

\newpage
\section{Additional data: Doping dependence}

\begin{figure} [!htbp] 
	\includegraphics[width=0.9\textwidth, angle=0]{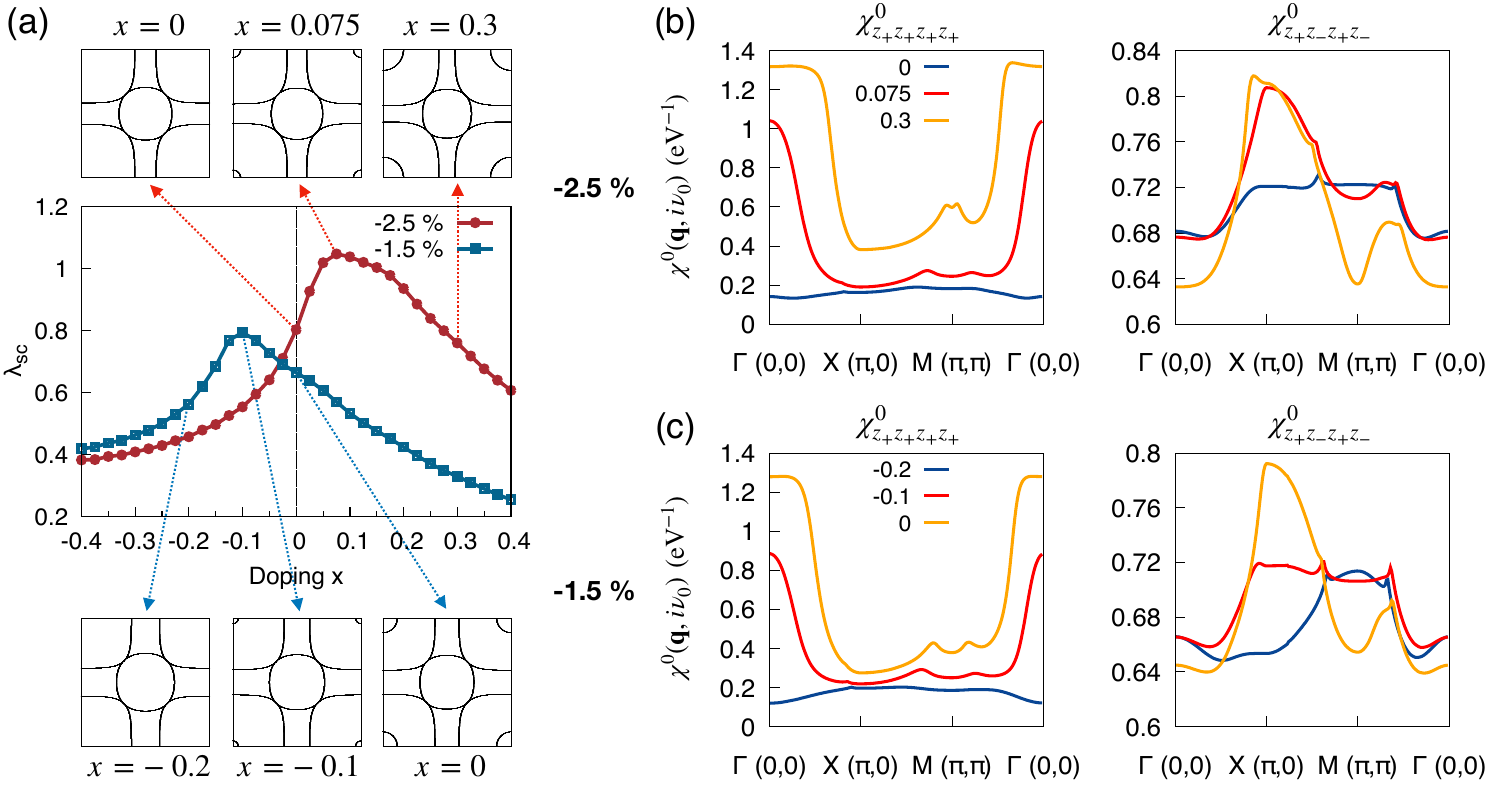}
	\caption{(a) The leading $\lambda_\mathrm{sc}$ ($s_\pm$-wave pairing) as a function of doping $x$ for $-2.5$~\% and $-1.5$~\% strain, and the corresponding FSs for selected cases. $x < 0$ corresponds to electron doping and $x > 0$ to hole doping.
    (b,c) The pair-breaking susceptibility $\chi^0_{z_+ z_+ z_+ z_+ }(\bm{q},i\nu_0)$ (left) and pair-forming susceptibility $\chi^0_{z_+ z_- z_+ z_- }(\bm{q},i\nu_0)$ (right) for several hole doping levels at (b) $-2.5$~\% strain and (c) $-1.5$~\% strain. $U=0.96$~eV, $J=U/6$, and $V=0$. }
	\label{sfig_doping}
\end{figure}

Figure~\ref{sfig_doping} shows the variation of the leading eigenvalue $\lambda_\mathrm{sc}$ as a function of doping $x$, where $x < 0$ corresponds to electron doping and $x > 0$ to hole doping. Notably, $\lambda_\mathrm{sc}$ reaches its maximum when the $\gamma$ pocket is marginally present in the Fermi surface for both $-2.5$~\% and $-1.5$~\% strain [Fig.~\ref{sfig_doping}(a)]. A closer inspection reveals that this behavior originates from a delicate balance between the pair-breaking channel ($\chi^0_{z_+ z_+ z_+ z_+}$) and the pair-forming channel ($\chi^0_{z_+ z_- z_+ z_-}$), both of which are promoted by the presence of the $\gamma$ pocket [Figs.~\ref{sfig_doping}(b,c)]. This interplay is analogous to what we have observed in the strain-dependent analysis throughout the main text.

%\newpage
%\vspace{in}
%\bibliographystyle{apsrev4-2}
%\bibliography{ref}
